\documentclass[reprint,superscriptaddress]{revtex4-1}
\usepackage{amsmath,bm,amssymb,amsfonts}
\usepackage{dcolumn}
\usepackage{graphicx}
\usepackage{latexsym}
\usepackage{epsfig}
\usepackage{multirow}
\usepackage{color}
\usepackage{bibunits}
\newcommand\scalemath[2]{\scalebox{#1}{\mbox{\ensuremath{\displaystyle #2}}}}

\begin{document}


\title{Strongly enhanced Rashba splittings in oxide heterostructure: a tantalate monolayer on BaHfO$_3$}

\author{Minsung \surname{Kim}}
\affiliation{Department of Physics and Astronomy, Seoul National
University, Seoul 151-747, Korea}
\affiliation{Ames Laboratory, US DOE and Department of Physics and Astronomy, Iowa State University, Ames, 
Iowa 50011, USA}
\author{Jisoon \surname{Ihm}}
 \email{To whom correspondence may be addressed. E-mail: jihm@snu.ac.kr, sbchung@snu.ac.kr}
\affiliation{Department of Physics and Astronomy, Seoul National University, Seoul 151-747, Korea}
\author{Suk Bum \surname{Chung}}
 \email{To whom correspondence may be addressed. E-mail: jihm@snu.ac.kr, sbchung@snu.ac.kr}
\affiliation{Department of Physics and Astronomy, Seoul National University, Seoul 151-747, Korea}
\affiliation{Center for Correlated Electron Systems, Institute for Basic Science (IBS), Seoul 151-747, Korea}

\date{\today}

\begin{abstract}
\textbf{
In the two-dimensional electron gas (2DEG) emerging at the transition metal oxide surface and interface, it has been pointed out that the Rashba spin-orbit interaction, the momentum-dependent spin splitting due to broken inversion symmetry and atomic spin-orbit coupling, can have profound effects on electronic ordering in the spin, orbit, and charge channels, and may help give rise to exotic phenomena such as ferromagnetism-superconductivity coexistence and topological superconductivity. Although a large Rashba splitting is expected to improve experimental accessibility of such phenomena, it has not been understood how we can maximally enhance this splitting. Here, we present a promising route to realize significant Rashba-type band splitting using a thin film heterostructure. Based on first-principles methods and analytic model analyses, a tantalate monolayer on BaHfO$_3$ is shown to host two-dimensional bands originating from Ta $t_{2g}$ states with strong Rashba spin splittings - up to nearly 10\% of the bandwidth - at both the band minima and saddle points due to the maximal breaking of the inversion symmetry. Such 2DEG band structure makes this oxide heterostructure a promising platform for realizing both a topological superconductor which hosts Majorana fermions and the electron correlation physics with strong spin-orbit coupling.
}
\end{abstract}


\date{\today}

\maketitle

Recently, the spin-orbit interaction of the two-dimensional electron gas (2DEG) at the surfaces and interfaces of the perovskite transition metal (TM) oxide
~\cite*{Ohtomo2004,Takagi2010,Mannhart2010,Santander2011,Meevasana2011} 
has been much investigated experimentally~\cite*{BenShalom2010,Caviglia2010,Santander2014,Santander2012,King2012,Reyren2012,MKim2012}. 
However, definite understanding on how its magnitude might be maximized has not been well established. 
It is the combination of the broken inversion symmetry and the atomic spin-orbit coupling (SOC) of the TM 
that gives rise to a non-zero spin splitting in the form of the 
Rashba spin-orbit interaction~\cite*{Bychkov1984,Dresselhaus1955,Winkler2003}. But this origin implies that the magnitude of the spin-orbit interaction is intrinsically limited by the TM atomic SOC. The limitation should be apparent in the best-studied perovskite 2DEGs --- the SrTiO$_3$ (STO) surface and the LaAlO$_3$/SrTiO$_3$ (LAO/STO) heterostructure interface ---
as the atomic SOC strength of the 3$d$ TM Ti is relatively small~\cite*{Zhong2013,Khalsa2013,PKim2014}.
Experimental evidences have been mixed, with the claims of large magnitude stemming from the magnetoresistance measurements
~\cite*{BenShalom2010,Caviglia2010,Santander2012,King2012,Santander2014} contradicted by the Hanle effect measurement~\cite*{Reyren2012} as well as the measurement of similar magnetoresistance in the $\delta$-doped STO heterostructure where the inversion symmetry breaking is hardly present~\cite*{MKim2012}. 
Meanwhile, theoretical calculations show 
the splitting near the $\Gamma$ point 
to be two orders of magnitude smaller than the bandwidth at the best ~\cite*{Zhong2013,YKim2013,PKim2014}. One natural way to overcome this limitation is adopting 5$d$ TM oxides, such as tantalate, with a stronger atomic SOC. This has motivated the recent experiments 
on the 2DEG at the surface of KTaO$_3$ (KTO)~\cite*{Santander2012,King2012}.  

However, the experiments on KTO have suggested that another important condition for enhancing the surface 2DEG 
Rashba spin-orbit interaction is to have the density profile of the surface state concentrated to the surface-terminating layer, which maximizes the effect of the broken inversion symmetry.
The ARPES measurements on the KTO surface have seen no measurable spin splitting~\cite*{Santander2012,King2012}, in spite of not only the stronger SOC of Ta but also the polar nature of KTO (001) surface. 
According to a density functional theory calculation~\cite*{Shanavas2014}, 
the surface state penetrates deeply into the bulk as the surface confinement potential is made shallow by the atomic relaxation near the surface layer. This suppresses the effect of the inversion symmetry breaking (ISB) on the surface state 
(which can be quantified by various parameters, {\it e.g.} the chiral orbital angular momentum coefficient~\cite*{SPark2011,BKim2012,CPark2012,PKim2014}),
and hence significantly reduces the 
Rashba spin-orbit interaction. 

\begin{figure}[]
\includegraphics[width=0.40\textwidth]{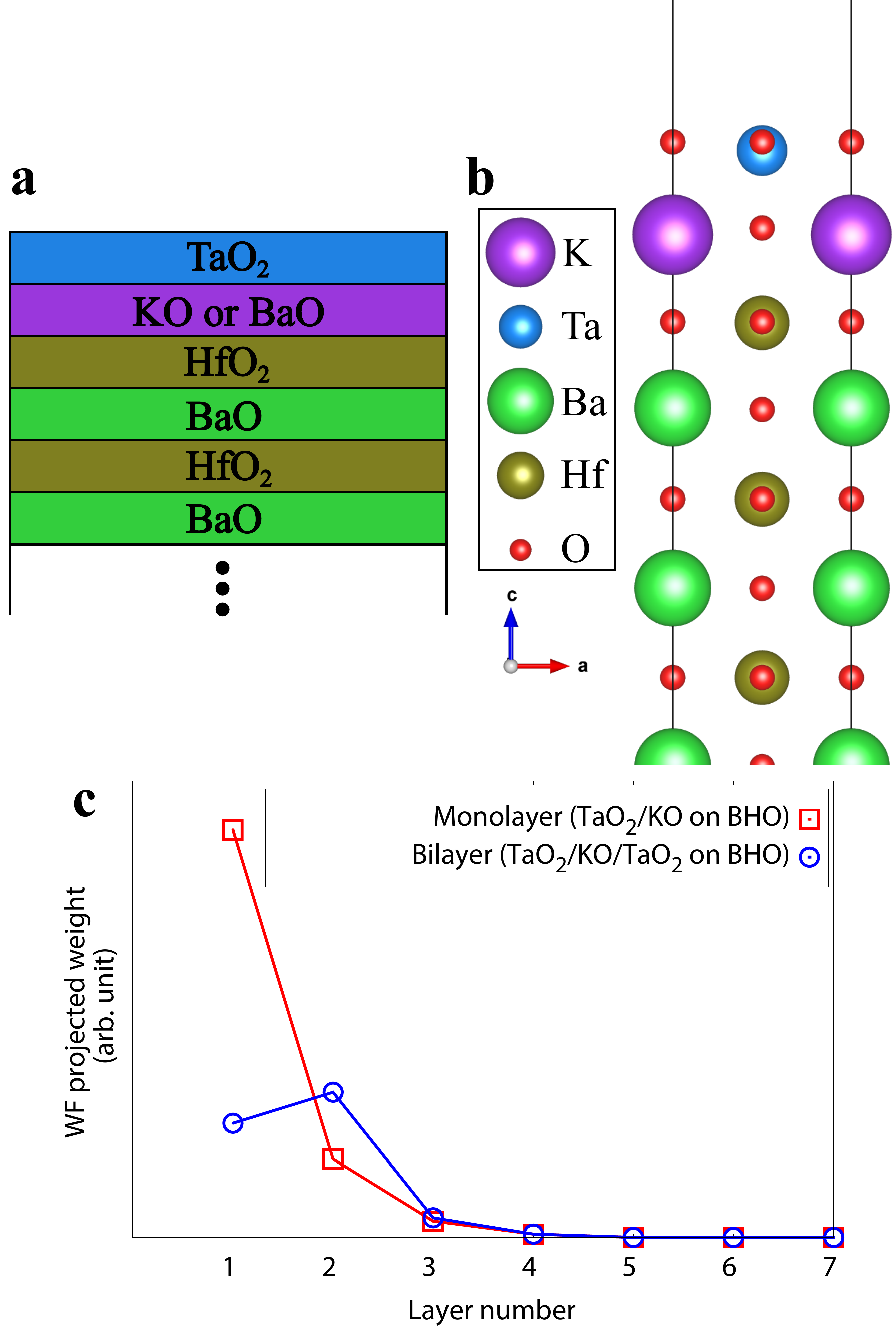}
\caption{\label{fig:atomicstr}Atomic structure of a tantalate layer on BaHfO$_3$ (001).
(a) Schematic illustration of TaO$_2$/KO on HfO$_2$-terminated BaHfO$_3$, and 
TaO$_2$ on BaO-terminated BaHfO$_3$. 
(b) Atomic structure of TaO$_2$/KO on BaHfO$_3$ from first-principles calculations; note the height difference between Ta and O at the top layer. 
(c) Wave function weight projected on $d_{xz/yz}$ of TMs 
for the lowest $d_{xz/yz}$ Rashba band at $\Gamma$.
Monolayer (TaO$_2$/KO on HfO$_2$-terminated BaHfO$_3$) case is to be compared with
bilayer (TaO$_2$/KO/TaO$_2$ on BaO-terminated BaHfO$_3$) case.
The TM-O$_2$ layers are numbered starting from the outermost layer.
}
\end{figure}

\begin{figure}[]
\includegraphics[width=0.45\textwidth]{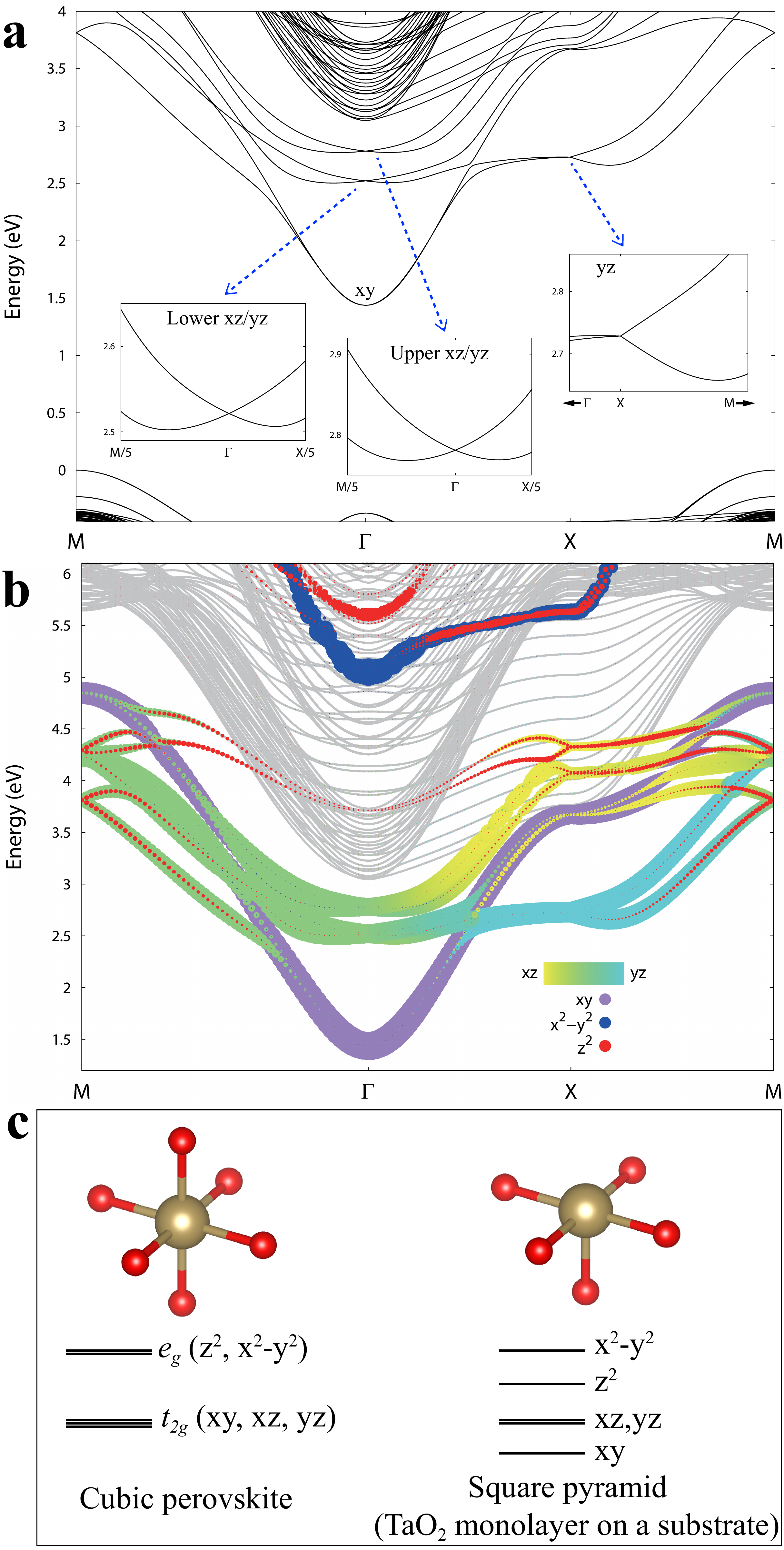}
\caption{\label{fig:bandpdos}Electronic structure of TaO$_2$/KO on BaHfO$_3$ from first-principles calculations. 
(a) Calculated band structure along high symmetry points. Insets show the magnified view  
of the upper and lower $d_{xz/yz}$ bands near $\Gamma$, 
and the $d_{yz}$ bands near $\mathrm{X}$.
(b) Projected weights of Ta $d$ states. The Fermi level is set to the valence band maximum.
(c) Schematic illustration of energy levels at $\Gamma$ without SOC.
The crystal field splitting of the monolayer-substrate heterostructure is different from that of the cubic (octahedral) case.}
\end{figure}

In this study, we have theoretically constructed a realistic oxide heterostructure that has a surface 2DEG  with a strong 
Rashba spin-orbit interaction. 
Our idea is to consider a 5$d$ TM oxide monolayer on a substrate, where 2DEG predominantly lies in the outermost monolayer film, maximizing the effect of the broken inversion symmetry from the substrate. 
Specifically, we attempt to replace the outer layers of perovskite oxide (001) surface with another perovskite thin film layers, in which the electronic bands of the substrate need to lie sufficiently far from the conduction band minimum (CBM) 
to make all essential low-energy physics originate from the thin film states near CBM.
After calculations of a number of candidate perovskite oxides for the heterostructure, we find that TaO$_2$/KO or TaO$_2$/BaO layer on BaHfO$_3$ (001) surface (Fig.~\ref{fig:atomicstr}) is 
a promising candidate structure possessing Ta $t_{2g}$-originated two-dimensional (2D) bands with 
a strong 
Rashba spin-orbit interaction. 
BaHfO$_3$ (BHO) is suitable as a substrate because its lattice structure (cubic at room temperature) matches with that of KTO, the only stable perovskite material containing TaO$_2$~\cite*{Zhurova2000,Maekawa2006}, and the alignment of its conduction bands and the Ta $t_{2g}$ bands enables minimal hybridization. 
It should be emphasized that the concentration of the surface state in the outermost layer is an important condition to maximize the ISB effect.
For instance, if we consider a TaO$_{2}$ bilayer as opposed to the monolayer (Fig.~\ref{fig:atomicstr}c), the ISB effect is weakened as the surface state wave function no longer peaks at the outermost layer. 
We will show that the coupling of the $t_{2g}$ surface bands to the Ta $e_g$ bands, 
which comes from the local asymmetric environment of the surface Ta atoms, plays a key role in the enhanced splitting. 
The band splitting from intra-$t_{2g}$ coupling is smaller, due to 
specific orbital symmetry of Ta $d$ and O $p$ states 
that we will discuss. 
We will further show that the $t_{2g}$-$e_g$ coupling gives rise to the enhanced Rashba-Dresselhaus splitting not only at the band bottom at $\Gamma$ but also at the band saddle points at $\mathrm{X}$. 
Finally, we will consider the substitution of Ba atoms for K in the KO layer for electron doping of the system.

\textbf{\large Results}

{\bf Rashba splitting near the $\Gamma$ point.}
The electronic structures of TaO$_2$/KO monolayer on BHO (lattice constant $\approx 4.155$ \AA) from our first-principles calculations are presented in Fig.~\ref{fig:bandpdos}. The bands near the CBM consist of $t_{2g}$ ($d_{xy}$, $d_{xz}$, $d_{yz}$) states of Ta in the outermost layer, with the calculated bandwidth of $\approx$ 1.7 eV for the $d_{xz/yz}$ bands. 
These bands being 2D, the triple degeneracy (excluding spin) of the $t_{2g}$ bands at $\Gamma$ is lifted, splitting the $d_{xy}$ and the $d_{xz/yz}$ manifolds; the Ta atomic SOC further splits the $d_{xz/yz}$ bands into upper and lower $d_{xz/yz}$ states. 
Finally, when the ISB at the surface is accounted for, the Rasha-type band splitting lifts spin degeneracies in the entire Brillouin zone (BZ) except at the time-reversal invariant momenta $\Gamma$, $\mathrm{X}$ and $\mathrm{M}$. 
We note that this Rashba-type band splitting of 
the $d_{xz/yz}$ bands is 
strikingly larger in magnitude than that of the $d_{xy}$ bands, contrary to the prediction of the $t_{2g}$-only 
model~\cite*{Khalsa2013,YKim2013,Scheurer2015}.  
Moreover, our calculation gives the Rashba coefficient of the lower $d_{xz/yz}$ bands at $\Gamma$ 
of $\alpha_R \approx 0.3~\mathrm{eV\AA}$, which is an order of magnitude larger than that of LAO/STO heterostructure 
deduced from the experimental magnetoresistance data~\cite*{Caviglia2010}, and the Rashba energy of $E_R \gtrsim 15~\mathrm{meV}$; these values are also significantly larger than $\alpha_R \approx 0.1~\mathrm{eV\AA}$, $E_R \approx 1~\mathrm{meV}$ for the bilayer case of Fig.~\ref{fig:atomicstr}c. 
The Rashba-Dresselhaus effect along the BZ boundary is even more pronounced, with a giant splitting ($\approx 180$ meV), which is nearly twice the maximum reported value~\cite*{Santander2014} in the perovskite oxide 2DEG,  occurring near $\mathrm{X}$ along $k_{x/y} = \pi$.

\begin{table}[b]
\caption{\label{tab:heff}%
Splitting terms of the effective Hamiltonian $\mathcal{H}_{eff}$ for Ta $t_{2g}$ manifolds. 
$\vec{k}$ denotes the reference point of the effective Hamiltonian 
with $\Gamma=(0,0)$ and $\mathrm{X}=(\pi,0)$.
$\Delta (\tilde{\Delta})$ represents the energy difference between the states in the subscript at $\Gamma (\mathrm{X})$,
where $uxz/yz$ and $lxz/yz$ mean the upper $xz/yz$ and the lower $xz/yz$, respectively.
}
\begin{ruledtabular}
\begin{tabular}{cccccc}
$\vec{k}$ & Manifold & Splitting terms in $\mathcal{H}_{\mathrm{eff}}$ \\
\colrule
\multirow{3}{*}{$\Gamma$} & upper $d_{xz/yz}$ & 
$\left[\frac{-2\gamma_3 \xi}{\Delta_{uxz/yz,x^2-y^2}}+\frac{-2\gamma_1 \xi}{\Delta_{uxz/yz,xy}}\right] 
(\vec{\sigma}\times \vec{k})\cdot \hat{z}$ \\
& lower $d_{xz/yz}$ & $\frac{2\sqrt{3}\gamma_2 \xi}{\Delta_{lxz/yz,z^2}} (\vec{\sigma}\times \vec{k})\cdot \hat{z}$ \\
& $d_{xy}$ & $\frac{-2\gamma_1 \xi}{\Delta_{xy,uxz/yz}} (\vec{\sigma}\times \vec{k})\cdot \hat{z}$ \\
\colrule
{\textrm{$\mathrm{X}$}} & $d_{yz}$ & $\left[\frac{-2\sqrt{3}\gamma_2 \xi}{\tilde{\Delta}_{yz,z^2}}
+\frac{2\gamma_3 \xi}{\tilde{\Delta}_{yz,x^2-y^2}}\right] 
\sigma_x k_y -\frac{2\gamma_1 \xi}{\tilde{\Delta}_{yz,xy}}\sigma_y k_x$ \\
\end{tabular}
\end{ruledtabular}
\end{table}

An analysis that includes all Ta $d$-orbitals -- not only the $t_{2g}$ orbitals but also the $e_g$ orbitals -- is required in order to understand the two conspicuous features of 
Fig.~\ref{fig:bandpdos}, the discrepancy between the Rashba splitting of the $d_{xy}$ and the $d_{xz/yz}$ bands, and the giant band splitting along $k_{x/y} = \pi$. 
We employ an analytic TB model for a qualitative 
analysis and supplement it with quantitative results from maximally localized Wannier functions (MLWFs). 
In the TB model, we consider a Hamiltonian for all Ta $d$-orbitals, including the $e_g$ orbitals ($d_{z^2}$, $d_{x^2-y^2}$) in a square lattice~\cite*{Shanavas2014,Shanavas2014a}  to describe 
the TaO$_2$ layer 2D bands, 
\begin{eqnarray}
\mathcal{H}=\mathcal{H}_\mathrm{hop}+\mathcal{H}_\mathrm{SOC}+\mathcal{H}_\mathrm{E}+\mathcal{V}_\mathrm{sf},
\end{eqnarray}
where the first term $\mathcal{H}_\mathrm{hop}$ describes the nearest-neighbor hopping, 
and the second term $\mathcal{H}_\mathrm{SOC} = \xi {\bf L} \cdot {\bf S}$ is the atomic SOC, with $\xi \approx 0.26$ eV for Ta. The third term $\mathcal{H}_\mathrm{E}$ includes the additional hoppings 
that would have been forbidden if not for the ISB:
\begin{eqnarray}
 &\gamma_1&=\langle d_{xy}| \mathcal{H}_\mathrm{E} |d_{xz} \rangle_{\hat{y}} 
 =\langle d_{xy}| \mathcal{H}_\mathrm{E} |d_{yz} \rangle_{\hat{x}} \nonumber \\ 
 &\gamma_2&=\langle d_{xz}| \mathcal{H}_\mathrm{E} |d_{z^2} \rangle_{\hat{x}} 
 =\langle d_{yz}| \mathcal{H}_\mathrm{E} |d_{z^2} \rangle_{\hat{y}} \\
 &\gamma_3&=\langle d_{x^2-y^2}| \mathcal{H}_\mathrm{E} |d_{yz} \rangle_{\hat{y}} 
 =\langle d_{xz}| \mathcal{H}_\mathrm{E} |d_{x^2-y^2} \rangle_{\hat{x}} \nonumber,
\end{eqnarray}
in which the vectors in the subscripts denote the relative position of the two orbitals with the lattice constant set to 1 for convenience 
(these ISB hoppings play a role analogous to the chiral orbital angular momentum effect in the $p$-orbital 
bands~\cite*{SPark2011,BKim2012,CPark2012}).
Here, $\gamma_1$ is intra-$t_{2g}$ ISB hopping while $\gamma_2$ and $\gamma_3$ describe $t_{2g}$-$e_g$ ISB hoppings. The fourth term $\mathcal{V}_\mathrm{sf}$ describes the potential difference 
originating from the surface field.
By deriving the effective Hamiltonian $\mathcal{H}_{eff}$ that acts on each two-fold degenerate band in the weak SOC limit where 
$\mathcal{H}_\mathrm{hop}+\mathcal{V}_\mathrm{sf}$ is dominant over $\mathcal{H}_\mathrm{SOC}$, 
we obtain Rashba-type band splitting terms near $\Gamma$ and $\mathrm{X}$ as summarized in Table~\ref{tab:heff} (see Appendix A for details).
Table~\ref{tab:heff} shows the Rashba coupling to be 
linear in the ISB hopping $\gamma$ divided by the energy difference between two relevant states $\Delta$. 

\begin{figure}[]
\includegraphics[width=0.35\textwidth]{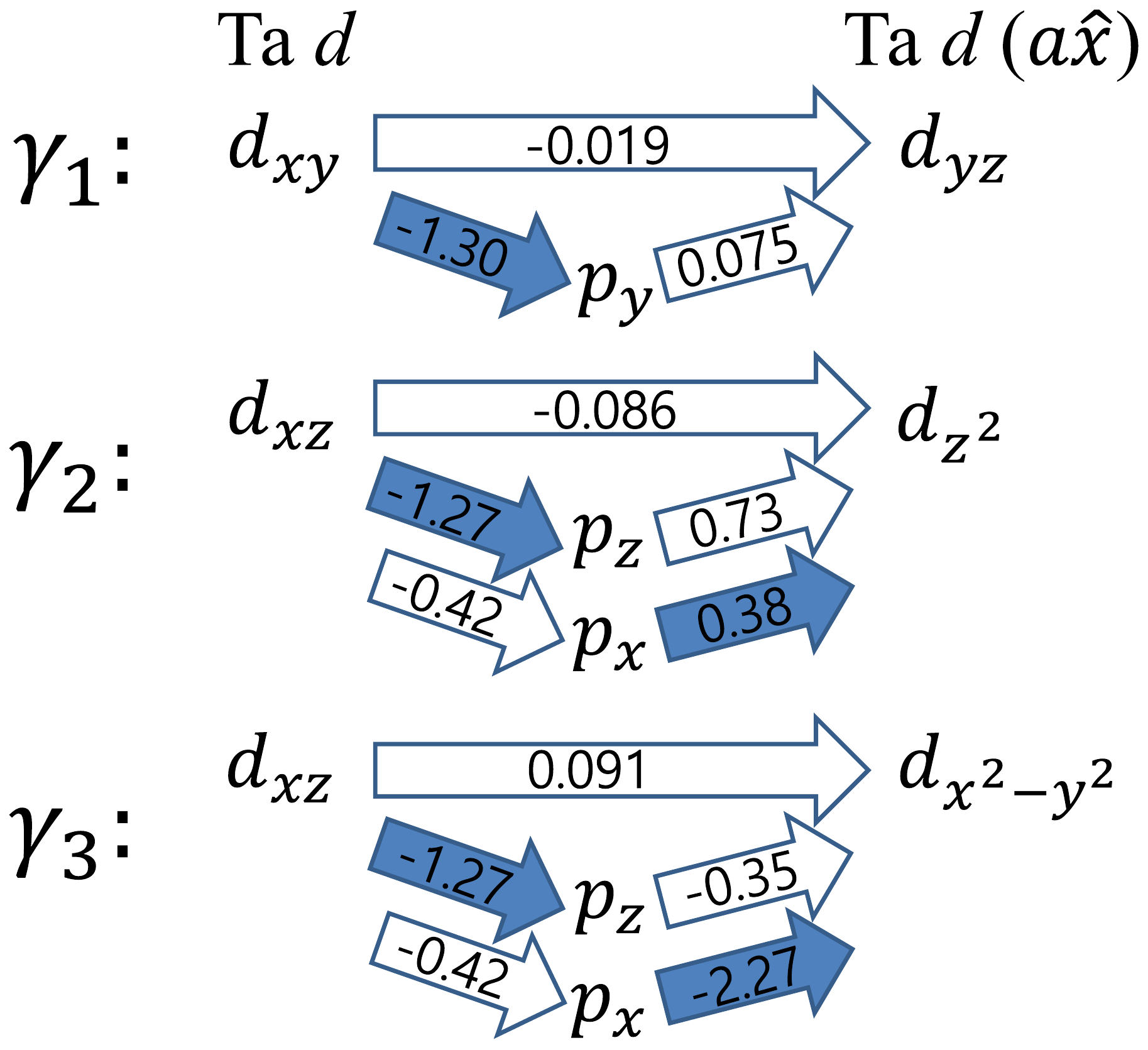}
\caption{\label{fig:wanhopping}Hopping strengths (in eV) between Wannier functions for Ta $d$ states at one site and the neighboring site in $x$ direction. 
Ta $d$ states and O $p$ states are used for the Wannier function construction.
The terms relevant to the ISB $\gamma_1$, $\gamma_2$, $\gamma_3$ are presented. 
Both direct (horizontal arrows) and indirect (oblique arrows, via O $p$) paths are depicted.
Empty arrows indicate terms that would be absent without ISB.}
\end{figure}

One reason why the $e_g$ orbital contribution is crucial for the Rashba splitting in the $t_{2g}$ bands is that the $t_{2g}$-$e_g$ ISB hoppings $\gamma_{2,3}$ are significantly larger than the intra-$t_{2g}$ hopping $\gamma_1$: $\gamma_1 \approx -0.04$ eV, $\gamma_2 \approx -0.25$ eV, $\gamma_3 \approx 0.30$ eV. This is necessary condition for the effective Hamiltonian of Table~\ref{tab:heff} to give larger Rashba splittings in the $d_{xz/yz}$ bands than $d_{xy}$ as shown in Fig.~\ref{fig:bandpdos}, given that the $d_{xz/yz}$ bands are closer in energy to the $d_{xy}$ band than the $e_g$ bands (albeit within an order of magnitude).
The intra-$t_{2g}$ ISB hopping $\gamma_1$ remains relatively small due to the orbital symmetry of Ta $d$ and O $p$ states,  
which we can see from a quantitative analysis with MLWFs that includes not only the Ta $d$ states but also the neighboring O $p$ states. 
Examining the hopping parameters relevant to $\gamma_1$, the 
particularly small ISB hopping between O $p_y$ and Ta $d_{yz}$ along $x$ direction 
(Fig.~\ref{fig:wanhopping}) 
can be attributed to the relative positions and shapes of the two orbitals; the lobes of the two orbitals lie on the $yz$ plane that is perpendicular to the hopping direction ($\hat{x}$), and $p_y$ has maximum amplitude along $y$ direction whereas $d_{yz}$ has a node along it. 
Thus, we have the negligible Rashba splitting of the $d_{xy}$ band as shown in Fig.~\ref{fig:bandpdos}, despite the smaller energy difference with the $d_{xz/yz}$ bands. 

The other reason why the $e_g$ orbital contribution is crucial for the Rashba splitting in the $t_{2g}$ bands is the reduced $t_{2g}$-$e_g$ energy splitting.
Indeed, when the $t_{2g}$-$e_g$ energy splitting is set to be infinite in Table~\ref{tab:heff}, all the results from the $t_{2g}$-only TB models~\cite*{Khalsa2013,YKim2013,PKim2014,Scheurer2015} are recovered, including the absence of $k$-linear Rashba in the lower $d_{xz/yz}$ band near $\Gamma$.  
In the case of 3D cubic KTO, the $t_{2g}$-$e_g$ energy separation at $\Gamma$ is calculated to be $\approx 4.6~\mathrm{eV}$,
which is larger than that of our system (Fig.~\ref{fig:bandpdos}a, b).
Compared with the 3D cubic bulk case, Figure~\ref{fig:bandpdos}b shows 
considerable portion of $d_{z^2}$ states 
close in energy, {\it i.e.}, less than bandwidth, to the $t_{2g}$ bands;   
this is due to the absence of an O atom in one of the octahedral points surrounding Ta. 
Hence, as shown in Fig.~\ref{fig:bandpdos}c, the local atomic configuration for the Ta atom is close to a square pyramid, where the $d_{z^2}$ and lower $d_{xz/yz}$ are close in energy. 
This can be taken as a generic result for the case where the 2DEG wave function is confined almost entirely to the outermost layer.
The height difference of Ta and O atoms ($\approx 0.20$ \AA) in the TaO$_2$ layer enhances the $t_{2g}$-$e_g$ coupling in both ways; the larger effect being the enhancement of the ISB hopping $\gamma_3$, but there is also noticeable lowering of the $d_{x^2-y^2}$ orbital energy level. 


\begin{figure}[]
\includegraphics[width=0.45\textwidth]{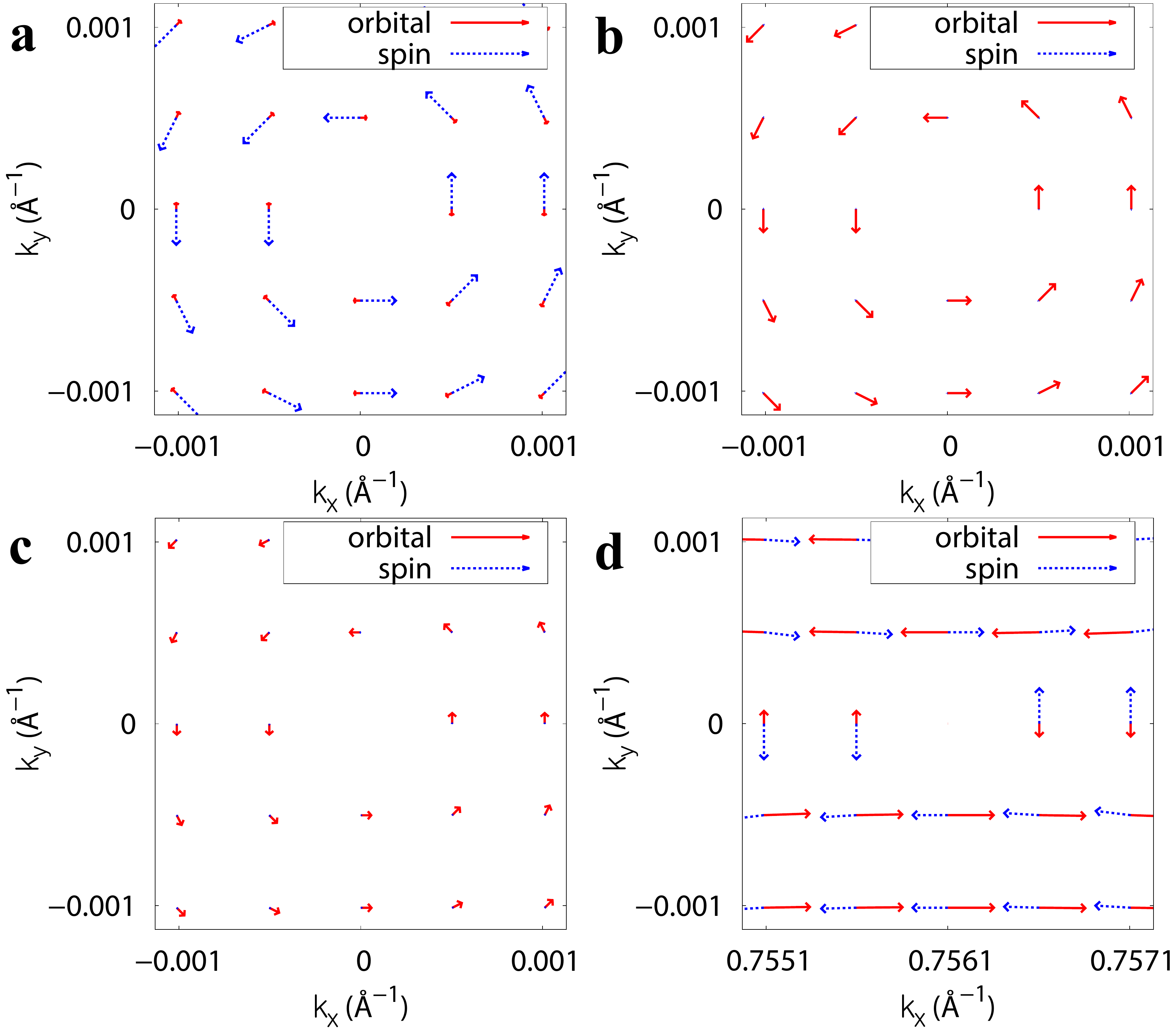}
\caption{\label{fig:am} Angular momentum texture from first-principles calculations 
in close vicinity of the high-symmetry points in BZ. 
Orbital and spin angular momenta are presented for the lower Rashba band of the (a) $d_{xy}$ and 
(b) lower $d_{xz/yz}$ (c) upper $d_{xz/yz}$ bands near $\Gamma$, 
and the lower Rashba-Dresselhaus band of the (d) $d_{yz}$ bands near $\mathrm{X}=(\pi,0)$.
}
\end{figure}

The $t_{2g}$-$e_g$ coupling also 
plays a key role in determining the angular momentum (AM) texture of the $t_{2g}$ bands in close vicinity of $\Gamma$ 
(Fig.~\ref{fig:am}). 
The tetragonal crystal field and the SOC determine the spin-orbital entanglement of the band manifolds; the spin-up and spin-down are in nearly the same orbital state for the $d_{xy}$ bands while they are in nearly orthogonal orbital states for the $d_{xz/yz}$ bands.
This, in turn, affects the AM character; the spin AM is dominant in the $d_{xy}$ bands whereas the orbital AM is dominant~\cite*{SPark2011,BKim2012,PKim2014} in the $d_{xz/yz}$ bands (see Appendix A).
The $t_{2g}$-$e_g$ coupling is important in that it gives rise to finite AM in the lower $d_{xz/yz}$ bands and quantitatively changes the AM in the upper $d_{xz/yz}$.
In the $t_{2g}$-only TB model, 
the AM texture of the lower $d_{xz/yz}$ band is completely missing
and that of the upper $d_{xz/yz}$ is not quantitatively correct. 

{\bf Band splitting near the $\mathrm{X}$ point.}
As shown in Table~\ref{tab:heff}, the lowered symmetry $C_{2v}$ at $\mathrm{X}$ allows the mixture of Rashba and linear Dresselhaus terms in general 
(see Appendix A),
with the linear Dresselhaus larger in magnitude, as shown in Fig.~\ref{fig:am}d. Due to the anisotropic dispersion of $d_{xz/yz}$ bands, 
the lowest conduction band at $\mathrm{X}=(\pi,0)$ mainly consists of $d_{yz}$ state. We find that the larger band splitting along $\mathrm{X}$---$\mathrm{M}$ comes from the $t_{2g}$-$e_g$ coupling whereas the 
smaller splitting along $\mathrm{X}$---$\Gamma$ is due to the intra-$t_{2g}$ coupling.
Hence, the 
giant Rashba-Dresselhaus splitting in vicinity of $\mathrm{X}$ ($\approx 180$ meV) is due to the $e_g$ contribution. 
It has been recently pointed out~\cite*{Chung2015} that this Rashba-Dresselhaus splitting along $\mathrm{X}$---$\mathrm{M}$ is necessary for weak topological superconductivity, which gives rise to dislocation Majorana zero modes.
The Rashba-Dresselhaus splitting near $\mathrm{X}$ also affects the superconducting instability, 
as it splits the logarithmic van Hove singularity (VHS) of the $d_{xz/yz}$ band saddle point and shifts them away from $\mathrm{X}$ 
(see Appendix B for the logarithmic VHS splitting).
Given that the splitting results in the lower and upper Rashba-Dresselhaus bands reaching VHS separately, the shifted VHSs do not have spin degeneracy. 
While it has been long recognized that the logarithmic VHS at $\mathrm{X}$ 
enhances the superconducting instability in the spin-singlet channel~\cite*{Schulz1987, Dzyaloshinskii1987, Furukawa1998, Gonzalez2008, Nandkishore2012}, 
it was recently shown~\cite*{Meng2015, Yao2015, Cheng2015} that the logarithmic VHS away from $\mathrm{X}$ enhances the instability to the spin-triplet $p$-wave superconductivity. 
The physics at $\mathrm{X}$ should be experimentally accessible, as the VHS at $\mathrm{X}$ is not too far from the band bottom of the $d_{xz/yz}$ in energy ($\approx 0.23$ eV), and we will show in the next section how our heterostructure can be chemically doped all the way to the VHS at $\mathrm{X}$. 

\textbf{\large Discussions}

\begin{figure}[]
\includegraphics[width=0.40\textwidth]{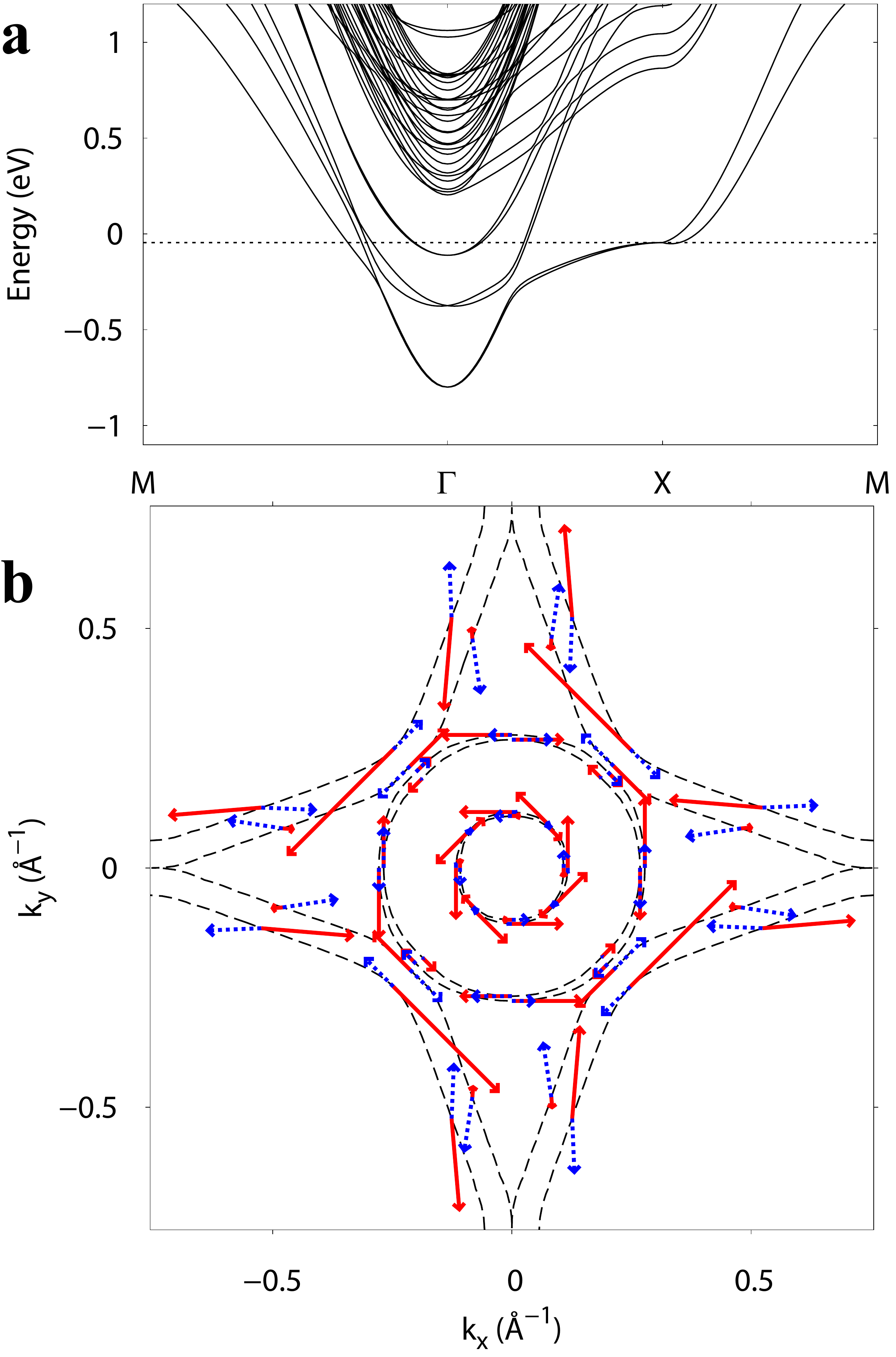}
\caption{\label{fig:tao2bho}Electronic structure of a TaO$_2$ layer on BaO-terminated
BaHfO$_3$ (001) from first-principles calculations.
(a) Band structure along high-symmetry points in BZ.
The dotted horizontal line denotes the energy level of the VHS in absence of the Rashba-Dresselhaus splitting, $E_{vH}$, 
with the Fermi level set to 0.
(b) AM texture at $E_{vH}$ with constant energy lines.
The red and blue arrows correspond to the orbital and spin AM, respectively. 
}
\end{figure}

To actually realize the 2DEG in the TaO$_2$ layer, electron doping is needed because the nominal charge of the TaO$_2$ layer is +1 and that of the KO layer is -1 (BaO and HfO$_2$ layers are neutral.).  One possible way would be substituting K atoms with Ba in the KO layer.
In this case, the Rashba strength remains still large ($\alpha_R \approx 0.2$ eV\AA) in the lower $d_{xz/yz}$ bands 
(Fig.~\ref{fig:tao2bho}a), while the Rashba splitting in the upper $d_{xz/yz}$ bands is greatly suppressed largely because the Ba substitution reduces by 70\% the height difference of Ta and O atoms in the TaO$_2$ layer. Once again, the considerable band splitting along $\mathrm{X}$--$\mathrm{M}$  ($\approx 24$ meV) and AM texture at $E_{vH}$ in Fig~\ref{fig:tao2bho} are contrary to the prediction of the $t_{2g}$-only TB model (see Appendix C and Fig. A1), and hence demonstrate that inclusion of the $e_g$ manifold is essential in understanding the Rashba-Dresselhaus splitting. 
We expect that partial chemical substitution (TaO$_2$/K$_{1-x}$Ba$_{x}$O layer on BHO) could induce the 2DEG in the TaO$_2$ layer in experiments. 
At $x=1$, {\it i.e.}, 100\% Ba substitution, the Fermi level lies 
slightly above the VHS (Fig.~\ref{fig:tao2bho}a), 
which means we can access the VHS at $x \lesssim 1$.  
We expect the qualitative features of our Rashba-Dresselhaus splitting to be generic for the (001) perovskite transition metal oxide 2DEG with maximal ISB, where the 2DEG wave function profile is required to be concentrated on the surface-terminating TM-O$_2$ layer.
In such an environment, the effective crystal field on the 2DEG $d$ orbitals should be quite different from that of the cubic perovskite, with much lower energy level at least for the $d_{z^2}$ orbital. 
Even if we use alternative materials for our heterostructure ---  
viability of substituting BHO by BaSnO$_3$ or TaO$_2$ by WO$_2$ still remains to be investigated 
--- we expect a significant role of the transition metal $e_g$ orbitals if the Rashba-Dresselhaus splitting 
is comparably large. 
It is also found that compressive strain on the BHO substrate, which might be needed for the feasible
deposition of the thin film taking into account relatively large lattice mismatch between KTO and BHO, does not substantially affect the band splitting 
(see Appendix D and Fig. A2).
Considering that 2DEG in an artificial film-substrate system is realized experimentally
in SrVO$_3$ thin films on Nb-doped STO~\cite*{Yoshimatsu2011}, we expect that our system can be realized in experiments 
using the state-of-the-art layer-by-layer growth control of perovskite oxide thin films~\cite*{Thiel2006}.



\textbf{\large Methods}

{\bf Theoretical approach.}
We performed density functional theory calculations as implemented in VASP~\cite*{Kresse1993,Kresse1996}. 
Projector augmented-wave method was used~\cite*{Blochl1994}.
A plane-wave basis set with the cutoff energy 520 eV was employed, and
PBEsol (Perdew-Burke-Ernzerhof revised for solids) exchange-correlation functional was adoped~\cite*{Perdew2008}.
We used the lattice constant optimized in bulk calculations of BaHfO$_3$ 
and the internal atomic positions were fully relaxed until the force became less than 0.01 eV/\AA.
Details of the analytic tight-binding approximation and effective Hamiltonian description were presented 
in Appendix A. 
We employed maximally localized Wannier functions~\cite*{Marzari1997,Souza2001,Mostofi2008} to further analyze the results of 
the first-principles calculations.
The Wannier functions were constructed for $d$ orbitals of Ta in one set,
and $p$ orbitals of three neighboring O as well as $d$ orbitals of Ta in the other set.

\textbf{\large Acknowledgments}

 We thank Jung Hoon Han, Changyoung Kim, Choong-Hyun Kim, Hyeong-Do Kim, Minu Kim, Hyun-Woo Lee, Hosub Jin, Seung Ryong Park, Cai-Zhuang Wang, Hong Yao, and Jaejun Yu for fruitful discussions and comments.
 This work was supported by the NRF of the MSIP of the Korean government Grant No. 2006-0093853 (M.K., J.I.) and IBS-R009-Y1 (S.B.C.).
 Research at Ames laboratory (M.K.) was supported by the U.S. Department of Energy, Office of Basic Energy Sciences, Division of Materials Sciences and Engineering under Contract No. DE-AC02-07CH11358.
 Computations were performed through the support of the Korea Institute of Science and Technology Information (KISTI) 
 and the National Energy Research Scientific Computing Center (NERSC) in Berkeley, CA.

\textbf{\large Author contributions}

 All authors contributed extensively to the work presented in this paper.

\appendix

\begin{widetext}
\section{Tight-binding model for the TaO$_2$ film}

We consider the tight-binding Hamiltonian for a TaO$_2$ film ($d$-orbitals in a square lattice)~\cite*{Shanavas2014a},
\begin{eqnarray}
 \mathcal{H}=\mathcal{H}_\mathrm{hop}+\mathcal{H}_\mathrm{SOC}+\mathcal{H}_\mathrm{E}+\mathcal{V}_\mathrm{sf},
\end{eqnarray}
where $\mathcal{H}_\mathrm{hop}$ describes the hopping between the nearest neighbors,
$\mathcal{H}_\mathrm{SOC}$ is the atomic spin-orbit coupling of Ta,
$\mathcal{H}_\mathrm{E}$ describes the orbital mixing due to the inversion symmetry breaking field near the surface,
and $\mathcal{V}_\mathrm{sf}$ describes onsite potential changes due to the surface field.
Specifically, the hopping term is given by
\begin{eqnarray}
 \mathcal{H}_\mathrm{hop}= 
 {
 \left( \begin{matrix}
 \frac{t_{\sigma}+3t_{\delta}}{2}(c_x+c_y) & -\frac{\sqrt{3}}{2}(t_{\sigma}-t_{\delta})(c_x-c_y) & 0 & 0 & 0 \\

 -\frac{\sqrt{3}}{2}(t_{\sigma}-t_{\delta})(c_x-c_y) & \frac{3t_{\sigma}+t_{\delta}}{2}(c_x+c_y) & 0 & 0 & 0 \\

 0 & 0 & 2t_{\pi}(c_x+c_y) & 0 & 0 \\

 0 & 0 & 0 & 2(t_{\pi} c_x +t_{\delta} c_y) & 0 \\

 0 & 0 & 0 & 0 & 2(t_{\delta} c_x +t_{\pi} c_y) \\
 \end{matrix} \right),
 }
\label{eq:ham_hopping}
\nonumber
\end{eqnarray}
where the basis is $\{ |d_{z^2}\rangle, |d_{x^2-y^2}\rangle, |d_{xy}\rangle, |d_{xz}\rangle, |d_{yz}\rangle \}$,
and $t_{\sigma}$, $t_{\pi}$, $t_{\delta}$ are hopping parameters between $d$-orbitals. $c_x$ means $\cos k_x$. 
The lattice constant is set to 1.
The spin-orbit coupling term is
\begin{eqnarray}
 \mathcal{H}_\mathrm{SOC}= 
 {
 \left( \begin{matrix}
 0 & 0 & 0 & 0 & 0 & 0 & 0 & -\frac{\sqrt{3}}{2}\xi & 0 & \frac{\sqrt{3}}{2}\xi i \\

 0 & 0 & 0 & 0 & 0 & 0 & \frac{\sqrt{3}}{2}\xi & 0 & \frac{\sqrt{3}}{2}\xi i & 0 \\

 0 & 0 & 0 & 0 & -\xi i & 0 & 0 & \frac{1}{2}\xi & 0 & \frac{1}{2}\xi i \\

 0 & 0 & 0 & 0 & 0 & \xi i & -\frac{1}{2}\xi & 0 & \frac{1}{2}\xi i & 0 \\

 0 & 0 & \xi i & 0 & 0 & 0 & 0 & -\frac{1}{2}\xi i & 0 & \frac{1}{2}\xi \\

 0 & 0 & 0 & -\xi i & 0 & 0 & -\frac{1}{2}\xi i & 0 & -\frac{1}{2}\xi & 0 \\ 

 0 & \frac{\sqrt{3}}{2}\xi & 0 & -\frac{1}{2}\xi 
 & 0 & \frac{1}{2}\xi i & 0 & 0 & -\frac{1}{2}\xi i & 0 \\

 -\frac{\sqrt{3}}{2}\xi & 0 & \frac{1}{2}\xi & 0 & \frac{1}{2}\xi i & 0 & 0 & 0 & 0 & \frac{1}{2}\xi i \\

 0 & -\frac{\sqrt{3}}{2}\xi i & 0 & -\frac{1}{2}\xi i & 0 & -\frac{1}{2}\xi & \frac{1}{2}\xi i & 0 & 0 & 0 \\

 -\frac{\sqrt{3}}{2}\xi i & 0 & -\frac{1}{2}\xi i & 0 & \frac{1}{2}\xi & 0 & 0 & -\frac{1}{2}\xi i & 0 & 0
  \end{matrix} \right)
 }
\label{eq:ham_soc}.
\nonumber
\end{eqnarray}
The inversion symmetry breaking field terms are given by
\begin{eqnarray}
 \mathcal{H}_\mathrm{E}+\mathcal{V}_\mathrm{sf}=  
 {
 \left( \begin{matrix}
 \delta_2 & 0 & 0 & -2i\gamma_2 \sin k_x & -2i\gamma_2 \sin k_y \\

 0 & \delta_3 & 0 & -2i\gamma_3 \sin k_x & 2i\gamma_3 \sin k_y \\

 0 & 0 & \delta_1 & 2i\gamma_1 \sin k_y & 2i\gamma_1 \sin k_x \\

 2i\gamma_2 \sin k_x & 2i\gamma_3 \sin k_x & -2i\gamma_1 \sin k_y & 0 & 0 \\

 2i\gamma_2 \sin k_y & -2i\gamma_3 \sin k_y & -2i\gamma_1 \sin k_x & 0 & 0 \\
 \end{matrix} \right),
 }
\label{eq:ham_efield}
\nonumber
\end{eqnarray}
where
\begin{eqnarray}
 &\delta_1&=\varepsilon(d_{xy})-\varepsilon(d_{xz/yz}) \\
 &\delta_2&=\varepsilon(d_{z^2})-\varepsilon(d_{xz/yz}) \\
 &\delta_3&=\varepsilon(d_{x^2-y^2})-\varepsilon(d_{xz/yz})\\
 &\gamma_1&=\langle d_{xy}| \mathcal{H}_\mathrm{E} |d_{xz} \rangle_{\hat{y}} 
 =\langle d_{xy}| \mathcal{H}_\mathrm{E} |d_{yz} \rangle_{\hat{x}} \\ 
 &\gamma_2&=\langle d_{xz}| \mathcal{H}_\mathrm{E} |d_{z^2} \rangle_{\hat{x}} 
 =\langle d_{yz}| \mathcal{H}_\mathrm{E} |d_{z^2} \rangle_{\hat{y}} \\
 &\gamma_3&=\langle d_{x^2-y^2}| \mathcal{H}_\mathrm{E} |d_{yz} \rangle_{\hat{y}} 
 =\langle d_{xz}| \mathcal{H}_\mathrm{E} |d_{x^2-y^2} \rangle_{\hat{x}} .
\end{eqnarray}

The Hamiltonian near the $\Gamma$ point can be written as
\begin{eqnarray*}
 \mathcal{H}(\vec{k})\approx \hspace{9cm} \\ 
 \scalemath{0.65}{
 \left( \begin{matrix}
 \frac{t_{\sigma}+3t_{\delta}}{2} C+\delta_2 & 0 & -\frac{\sqrt{3}}{2}(t_{\sigma}-t_{\delta}) D & 0
 & 0 & 0 & 0 & -\sqrt{2}\gamma_2 (i k_x-k_y) & -\sqrt{\frac{3}{2}} \xi & -\sqrt{2}\gamma_2 (i k_x+k_y) \\

 0 & \frac{t_{\sigma}+3t_{\delta}}{2} C+\delta_2 & 0 & -\frac{\sqrt{3}}{2}(t_{\sigma}-t_{\delta}) D 
 & 0 & 0 & -\sqrt{2}\gamma_2 (i k_x+k_y) & 0 & -\sqrt{2}\gamma_2 (i k_x-k_y) & \sqrt{\frac{3}{2}} \xi \\

 -\frac{\sqrt{3}}{2}(t_{\sigma}-t_{\delta}) D & 0 & \frac{3t_{\sigma}+t_{\delta}}{2} C+\delta_3 & 0 
 & -\xi i & 0 & \frac{1}{\sqrt{2}} \xi & -\sqrt{2}\gamma_3 (i k_x+k_y) & 0 & -\sqrt{2}\gamma_3 (i k_x-k_y) \\

 0 & -\frac{\sqrt{3}}{2}(t_{\sigma}-t_{\delta}) D & 0 & \frac{3t_{\sigma}+t_{\delta}}{2} C+\delta_3  
 & 0 & \xi i & -\sqrt{2}\gamma_3 (i k_x-k_y) & -\frac{1}{\sqrt{2}} \xi & -\sqrt{2}\gamma_3 (i k_x+k_y) & 0 \\

 0 & 0 & \xi i & 0 
 & 2t_{\pi} C+\delta_1 & 0 & -\frac{1}{\sqrt{2}} i \xi & \sqrt{2} i \gamma_1 (i k_x+k_y) & 0 & -\sqrt{2} i \gamma_1 (i k_x-k_y) \\

 0 & 0 & 0 & -\xi i 
 & 0 & 2t_{\pi} C+\delta_1 & -\sqrt{2} i \gamma_1 (i k_x-k_y) & -\frac{1}{\sqrt{2}} i \xi & \sqrt{2} i \gamma_1 (i k_x+k_y) & 0 \\ 

 0 & \sqrt{2}\gamma_2 (i k_x-k_y) & \frac{1}{\sqrt{2}} \xi & \sqrt{2}\gamma_3 (i k_x+k_y) 
 & \frac{1}{\sqrt{2}} i \xi & -\sqrt{2} i \gamma_1 (i k_x+k_y) & (t_{\pi} +t_{\delta}) C+\frac{\xi}{2} & 0 & (t_{\pi} -t_{\delta}) D & 0 \\

 \sqrt{2}\gamma_2 (i k_x+k_y) & 0 & \sqrt{2}\gamma_3 (i k_x-k_y) & -\frac{1}{\sqrt{2}} \xi  
 & \sqrt{2} i \gamma_1 (i k_x-k_y) & \frac{1}{\sqrt{2}} i \xi & 0 & (t_{\pi} +t_{\delta}) C+\frac{\xi}{2} & 0 & (t_{\pi} -t_{\delta}) D \\

 -\sqrt{\frac{3}{2}} \xi & \sqrt{2}\gamma_2 (i k_x+k_y) & 0 & \sqrt{2}\gamma_3 (i k_x-k_y) 
 & 0 & \sqrt{2} i \gamma_1 (i k_x-k_y) & (t_{\pi} -t_{\delta}) D & 0 & (t_{\pi} +t_{\delta}) C-\frac{\xi}{2} & 0 \\

 \sqrt{2}\gamma_2 (i k_x-k_y) & \sqrt{\frac{3}{2}} \xi & \sqrt{2}\gamma_3 (i k_x+k_y) & 0 
 & -\sqrt{2} i \gamma_1 (i k_x+k_y) & 0 & 0 & (t_{\pi} -t_{\delta}) D & 0 & (t_{\pi} +t_{\delta}) C-\frac{\xi}{2}
 \end{matrix} \right),
 }
\label{eq:ham_k_neargamma}
\nonumber
\end{eqnarray*}
with $C=\cos k_x + \cos k_y \approx 2-\frac{k_x^2}{2}-\frac{k_y^2}{2}$, 
$D=\cos k_x - \cos k_y \approx -\frac{k_x^2}{2}+\frac{k_y^2}{2}$,
where we performed a unitary transformation to diagonalize the $d_{xz/yz}$ subspace in the limit 
that the $d_{xz/yz}$ states are sufficiently far from other manifolds and $\vec{k} \rightarrow 0$.

The effective Hamiltonian can be obtained by projection onto the concerned manifold
\begin{eqnarray}
 \mathcal{H}_{\mathrm{eff}}=\mathcal{P}\mathcal{H}\mathcal{P}+\mathcal{P}\mathcal{H}\mathcal{Q}
 \frac{1}{\epsilon-\mathcal{Q}\mathcal{H}\mathcal{Q}}\mathcal{Q}\mathcal{H}\mathcal{P},
\label{eq:ham_eff}
\end{eqnarray}
where $\mathcal{P}$ is the projection operator onto the relevant manifold and $\mathcal{Q}=1-\mathcal{P}$.
For the $d_{xy}$ bands, the effective Hamiltonian is
\begin{eqnarray}
 \mathcal{H}_{\mathrm{eff}}
 \approx h_{xy}(\vec{k})I_{2\times 2}+\frac{-2\gamma_1 \xi}{\Delta_{xy,uxz/yz}}
 (\vec{\sigma}\times \vec{k})\cdot \hat{z},
\label{eq:ham_eff_gamma_dxy}
\end{eqnarray}
where $\Delta_{xy,uxz/yz}=4t_{\pi}+\delta_1-\{2(t_{\pi}+t_{\sigma})+\frac{\xi}{2}\}$, and
the Pauli matrices describe the subspace defined by 
$\{ |d_{xy}\uparrow\rangle, |d_{xy}\downarrow\rangle \}$.
For the lower $d_{xz/yz}$ bands, 
\begin{eqnarray}
 \mathcal{H}_{\mathrm{eff}}
 \approx h_{lxz/yz}(\vec{k})I_{2\times 2}
 +\frac{2\sqrt{3}\gamma_2 \xi}{\Delta_{lxz/yz,z^2}}(\vec{\sigma}\times \vec{k})\cdot \hat{z},
\label{eq:ham_eff_gamma_lxz/yz}
\end{eqnarray}
where $\Delta_{lxz/yz,z^2}=2(t_{\pi}+t_{\delta})-\frac{\xi}{2}-\{t_{\sigma}+3t_{\delta}+\delta_2\}$,
and the Pauli matrices describe the subspace defined by 
$\{ \frac{1}{\sqrt{2}}(|d_{xz}\downarrow\rangle +i |d_{yz}\downarrow\rangle), 
\frac{1}{\sqrt{2}}(|d_{xz}\uparrow\rangle -i |d_{yz}\uparrow\rangle) \}$.
For the upper $d_{xz/yz}$ bands,
\begin{eqnarray}
 \mathcal{H}_{\mathrm{eff}}
 \approx h_{uxz/yz}(\vec{k})I_{2\times 2}
 +\left[\frac{-2\gamma_3 \xi}{\Delta_{uxz/yz,x^2-y^2}}+\frac{-2\gamma_1 \xi}{\Delta_{uxz/yz,xy}}\right]
 (\vec{\sigma}\times \vec{k})\cdot \hat{z},
\label{eq:ham_eff_gamma_uxz/yz}
\end{eqnarray}
where $\Delta_{uxz/yz,x^2-y^2}=2(t_{\pi}+t_{\delta})+\frac{\xi}{2}-\{3t_{\sigma}+t_{\delta}+\delta_3\}$,
$\Delta_{uxz/yz,xy}=2(t_{\pi}+t_{\delta})+\frac{\xi}{2}-\{4t_{\pi}+\delta_1\}$, 
and the Pauli matrices describe the subspace defined by
$\{ \frac{1}{\sqrt{2}}(|d_{xz}\downarrow\rangle -i|d_{yz}\downarrow\rangle), 
\frac{1}{\sqrt{2}}(|d_{xz}\uparrow\rangle +i|d_{yz}\uparrow\rangle) \}$.

The angular momentum (AM) texture can be calculated using the eigenstates with the lowest perturbative correction in $\xi$. 
For the $d_{xy}$ manifold in close vicinity of the $\Gamma$ point, 
the dominant spin AM expectation value $\langle S_y \rangle \approx \hbar/2$ for an eigenstate in $x$ direction 
comes from the original $d_{xy}$ manifold.
The remnant orbital AM $\langle L_y \rangle \approx\hbar \xi/\Delta_{xy,uxz/yz}$ is due to the inter-band coupling to the upper $d_{xz/yz}$, which can be calculated using the eigenstate with the first-order correction in $\xi$ that hybridizes the $d_{xy}$ manifold with the upper $d_{xz/yz}$ and the $d_{x^2-y^2}$ manifolds. 
As for the lower  $d_{xz/yz}$ manifold, 
both the orbital and spin AM expectation values 
 \begin{eqnarray}
 \langle L_y \rangle &\approx& \frac{-3 \hbar \xi}{\Delta_{lxz/yz,z^2}} \\
 \langle S_y \rangle &\approx& -\frac{3}{4} \hbar \left( \frac{\xi}{\Delta_{lxz/yz,z^2}} \right)^2.
 \label{eq:ham_R_lxz/yz_correction_lysy}
 \end{eqnarray}
for an eigenstate in $x$ direction can be obtained only from the eigenstates with the first-order correction in $\xi$ which leads to hybridization with the $d_{z^2}$ manifold. 
Thus, the orbital dominant AM texture in the lower $d_{xz/yz}$ bands comes from the inter-band coupling to the $d_{z^2}$.
Similarly, we find that the orbital dominant AM texture in the upper $d_{xz/yz}$ bands originates from 
the inter-band coupling to 
the $d_{xy}$ and the $d_{x^2-y^2}$ manifolds.

Near the $\mathrm{X}=(\pi,0)$ point, the effective Hamiltonian is 
\begin{eqnarray*}
 \mathcal{H}_{\mathrm{eff}}
 &=&h_{yz}(\vec{k})I_{2\times 2}
 +\left[\frac{-2\sqrt{3}\gamma_2 \xi}{\tilde{\Delta}_{yz,z^2}}+\frac{2\gamma_3 \xi}{\tilde{\Delta}_{yz,x^2-y^2}}\right]
 \sigma_x k_y -\frac{2\gamma_1 \xi}{\tilde{\Delta}_{yz,xy}}\sigma_y k_x,
\label{eq:ham_eff_X_yz}
\end{eqnarray*}
where  
$\tilde{\Delta}_{yz,z^2}=2(t_{\pi}-t_{\delta})-\delta_2$,
$\tilde{\Delta}_{yz,x^2-y^2}=2(t_{\pi}-t_{\delta})-\delta_3$,
$\tilde{\Delta}_{yz,xy}=2(t_{\pi}-t_{\delta})-\delta_1$, and the Pauli matrices describe the subspace defined by
$\{ | d_{yz}\uparrow\rangle, | d_{yz}\downarrow\rangle \}$, and
$(k_x, k_y)$ is a local coordinate with respect to $(\pi,0)$.
Here, the splitting terms are mixture of Rashba and linear Dresselhaus terms, which are of the form
\begin{eqnarray*}
\mathcal{H}_{\mathrm{splitting}}=A\sigma_x k_y -B\sigma_y k_x,
\end{eqnarray*}
with $A=\frac{-2\sqrt{3}\gamma_2 \xi}{\tilde{\Delta}_{yz,z^2}}+\frac{2\gamma_3 \xi}{\tilde{\Delta}_{yz,x^2-y^2}}$ and
$B=\frac{2\gamma_1 \xi}{\tilde{\Delta}_{yz,xy}}$.
If we rotate the local coordinate by $\pi/4$ about $k_z$ axis, 
the splitting terms become
\begin{eqnarray*}
\mathcal{H}_{\mathrm{splitting}}&=&\frac{A+B}{2}(\sigma_x k_y -\sigma_y k_x)
+\frac{A-B}{2}(\sigma_x k_x -\sigma_y k_y) \\
&=&\alpha_R(\sigma_x k_y -\sigma_y k_x)
+\alpha_D(\sigma_x k_x -\sigma_y k_y).
\end{eqnarray*}
In our case, we have $|A|\gg|B|$, thus both Rashba and linear Dresselhaus terms are present with similar strength.
Due to the symmetry, only Rashba term is allowed for $C_{4v}$ at $\mathrm{\Gamma}$ 
(where we should have $A=B$), 
and both Rashba and linear Dresselhaus terms are allowed for $C_{2v}$ at $\mathrm{X}$
~\cite*{Stroppa2014}.

\renewcommand{\thefigure}{A\arabic{figure}}

\setcounter{figure}{0}

\begin{figure}[]
\includegraphics[width=0.90\textwidth]{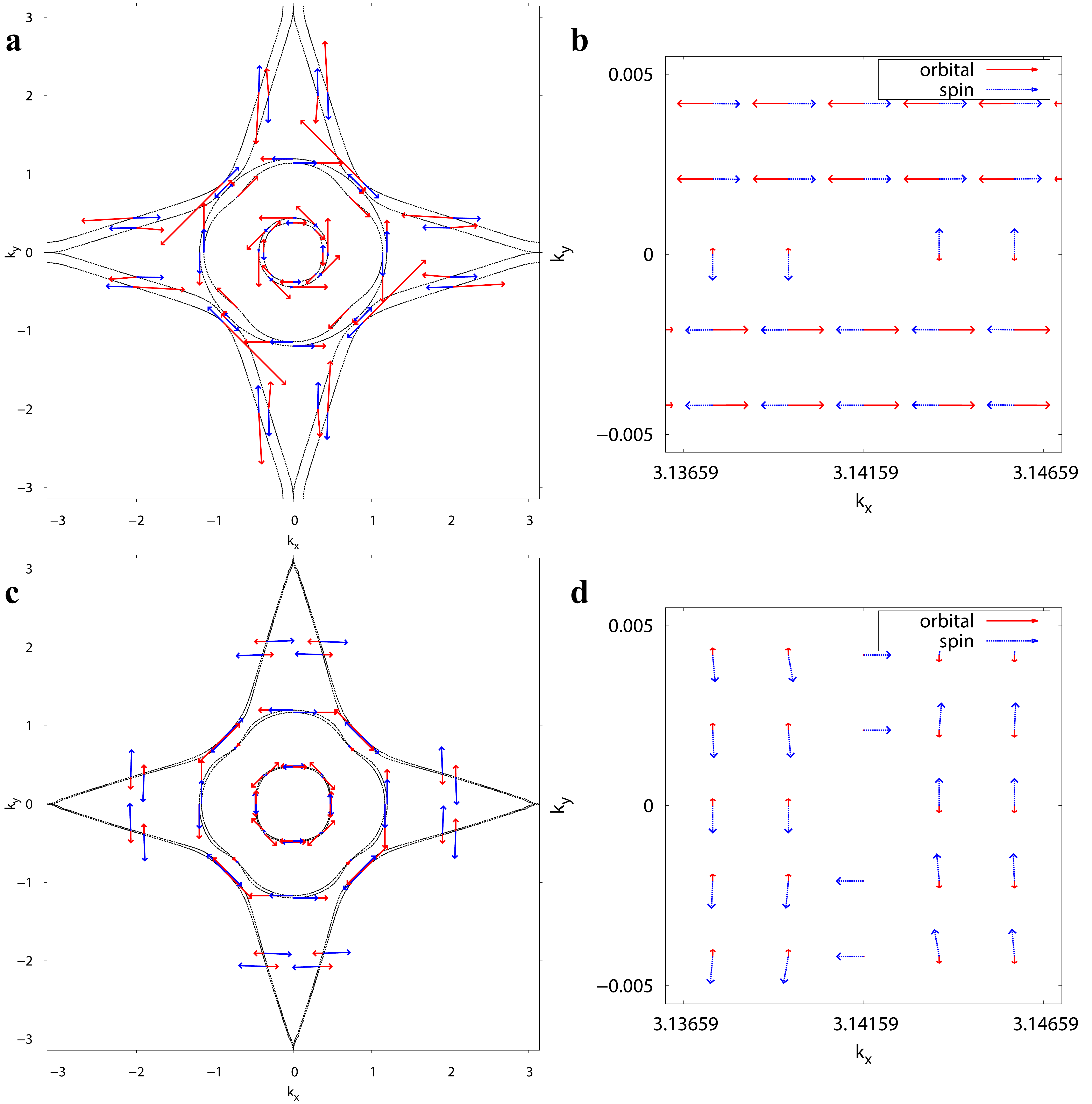}
\caption{\label{fig:tb_am} Angular momentum texture from the tight-binding model.
The angular momentum textures are calculated (a) at $E_{vH}$ and (b) near $\mathrm{X}$ using both $t_{2g}$ and $e_g$, 
and (c) at $E_{vH}$ and (d) near $\mathrm{X}$ using only $t_{2g}$.
The red and blue arrows represent the orbital and spin AM, respectively.
}
\end{figure}

\end{widetext}

\section{The splitting of the log van Hove singularity at $\mathrm{X}$}

We show here that the Rashba-Dresselhaus splitting removes the spin degeneracy of the logarithmic van Hove singularity at $\mathrm{X}$, resulting in the two separate logarithmic van Hove singularities for the upper and lower Rashba-Dresselhaus bands. This implies that there will be a very large density of state change between the upper and lower Rashba-Dresselhaus bands, which would have a significant effect on the phase competition, {\it e.g.} the relative magnitude of the pairing susceptibilities with different symmetries. 
 
It is well-known that there is a logarithmic van Hove singularity at $\mathrm{X}$ in absence of the Rashba-Dresselhaus splitting. 
The dispersion of the lowest energy band near $\mathrm{X}=(\pi,0)$ approximately follows the dispersion of the $d_{yz}$ band, 
$$\xi = 2(t_\delta \cos k_x + t_\pi \cos k_y) \approx t_\delta (k_x -\pi)^2 - t_\pi k_y^2 + 2(t_\pi - t_\delta);$$
it is well-understood that there is a logarithmic van Hove singularity at the saddle point of a quadratic Hamiltonian in 2D
~\cite*{Yu2010}.

The addition of the Rashba-Dresselhaus term near $\mathrm{X}$, $\mathcal{H}_{R-D} = A\sigma_x k_y - B \sigma_y (k_x - \pi)$, leads to the spin splitting of this saddle point, which modifies the dispersion to
\begin{equation*}
\xi_\pm \approx t_\delta (k_x -\pi)^2 - t_\pi k_y^2  \pm \sqrt{A^2 k_y^2 + B^2 (k_x -\pi)^2} + 2(t_\pi - t_\delta).
\end{equation*} 
Using the fact that the Fermi velocity vanishes when the van Hove singularity occurs, we can see that the van Hove singularity at $\mathrm{X} = (\pi, 0)$ is shifted to $(\pi \pm B/2t_\delta, 0)$ for the upper Rashba-Dresselhaus band, with the dispersion in its vicinity
$$\xi_+ \approx t_\delta \left(k_x - \pi \mp \frac{B}{2t_\delta}\right)^2 - \left(t_\pi + t_\delta \frac{A^2}{B^2}\right) k_y^2
+ 2(t_\pi - t_\delta) - \frac{B^2}{4t_\delta}$$
and $(\pi, \pm A/2t_\pi)$ for the lower Rashba-Dresselhaus band, with the dispersion in its vicinity
$$ \xi_- \approx \left(t_\delta + t_\pi \frac{B^2}{A^2}\right)(k_x - \pi)^2 - t_\pi (k_y \mp \frac{A}{2t_\pi})^2 + 2(t_\pi - t_\delta) + \frac{A^2}{4t_\pi}.$$

We see here that when we raise the chemical potential so that the Fermi surface passes through the ${\rm X}$ point, 
the Fermi level first passes through the logarithmic van Hove singularity of the lower Rashba-Dresselhaus band, and then that of the upper Rashba-Dresselhaus band.

\section{The importance of $e_g$ manifold in the angular momentum texture}

Because the $e_g$ manifold affects the Rashba-Dresselhaus splitting, 
the inclusion of the $e_g$ manifold is important to correctly describe the AM texture.
By numerically solving the tight-binding model, we obtained the AM expectation values with and without $e_g$ manifold
(Fig.~\ref{fig:tb_am}).
For the $t_{2g}$-only limit, we set $\delta_2 \approx \delta_3 \approx 10^3 \mathrm{eV}$.
We find considerable differences in view of the direction and magnitude of the AM.
Notably, the coupling to $e_g$ manifold has significant effects in the direction of the AM near $\mathrm{X}$ and in the intermediate region.

\section{The effect of compressive strain in the substrate}

Due to the large lattice constant of BaHfO$_3$, it might be helpful to apply compressive strain to the substrate
for the deposition of the tantalate thin film. 
The electronic band structure of TaO$_2$/KO layer on HfO$_2$-terminated BaHfO$_3$ with the lattice constant reduced by 2\%
is presented in Fig.~\ref{fig:strain_band}. 
We find that the Rashba coefficient remains still large 
(for example, $\alpha_R \approx 0.3~\mathrm{eV\AA}$ in the lower $d_{xz/yz}$ bands at $\Gamma$). 

\section{The relation between the band effective mass and Rashba-related parameters}

Here, we show that both the momentum offset $k_R$ and the Rashba energy $E_R$ are proportional to the effective mass
of the Rashba bands for a given Rashba strength $\alpha_R$.
We consider the Hamiltonian
\begin{eqnarray*}
 \mathcal{H}
 =\frac{\hbar^2 k^2}{2m^*} I_{2\times2}+\alpha_R (\vec{\sigma}\times \vec{k})\cdot \hat{z},
\end{eqnarray*}
with $k=\sqrt{k_x^2+k_y^2}$, where $m^*$ is the effective mass of the band and $I_{2\times2}$ is the $2\times2$ identity matrix.
The energy dispersion of the lower Rashba band is given by
\begin{eqnarray*}
 E(k)
 &=&\frac{\hbar^2 k^2}{2m^*} - |\alpha_R| k \\
 &=&\frac{\hbar^2}{2m^*} (k-\frac{m^* |\alpha_R|}{\hbar^2})^2 - \frac{m^* |\alpha_R|^2}{2\hbar^2} \\
 &\equiv&\frac{\hbar^2}{2m^*} (k-k_R)^2 - E_R.
\end{eqnarray*}
We find that the momentum offset $k_R = \frac{m^* |\alpha_R|}{\hbar^2}$ and
the Rashba energy $E_R = \frac{m^* |\alpha_R|^2}{2\hbar^2}$, which are principal measures of the band splitting size
when one sees a band structure figure, are proportional to the effective mass $m^*$ 
for a given Rashba parameter $\alpha_R$.
Thus, the Rashba splitting of the $d_{xz/yz}$ bands would look more pronounced due to the heavier effective mass
compared with the $d_{xy}$ band even if they had the same Rashba strength.

\renewcommand{\thefigure}{A\arabic{figure}}

\begin{figure}[h]
\includegraphics[width=0.45\textwidth]{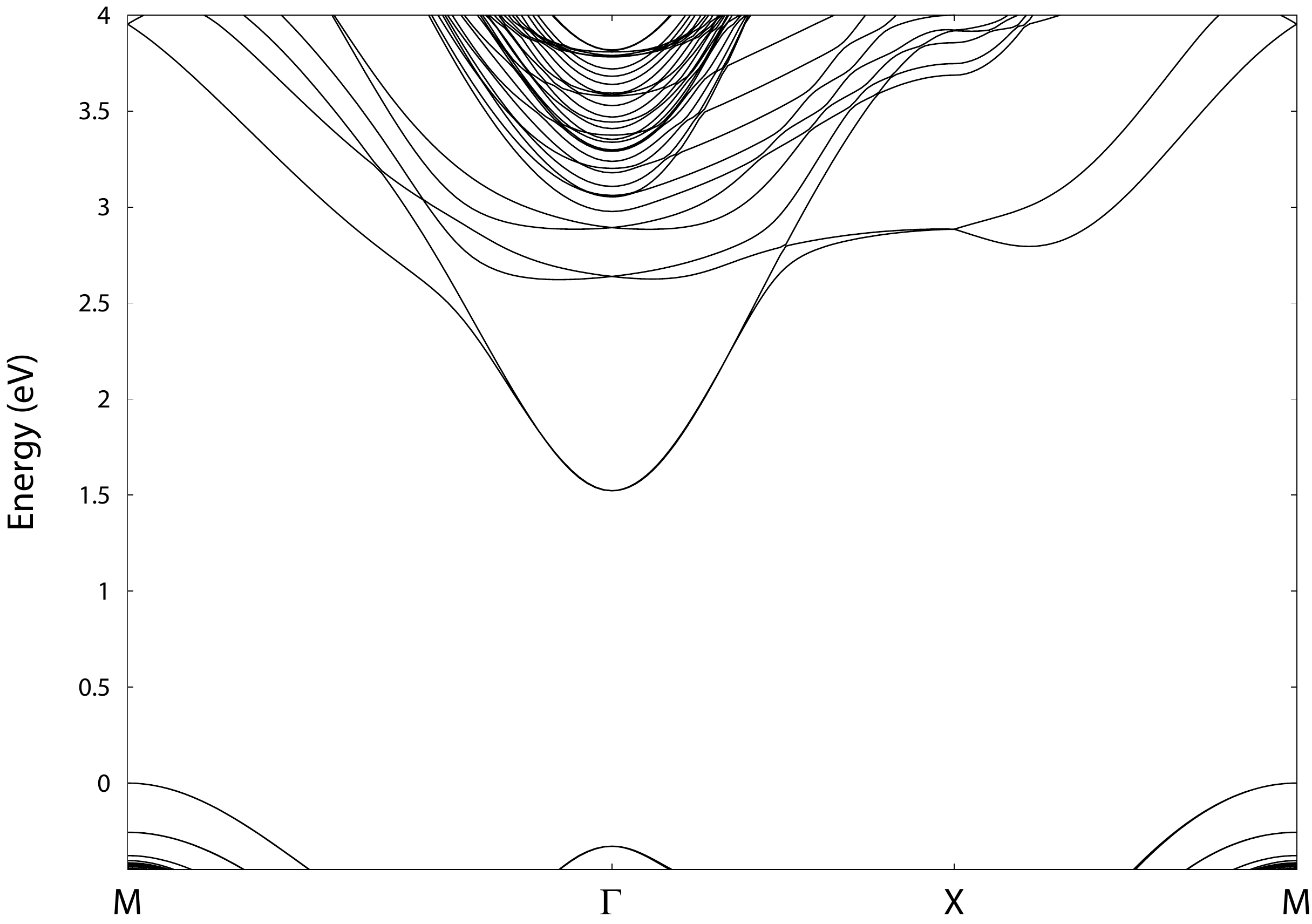}
\caption{\label{fig:strain_band} Band structure of TaO$_2$/KO  
on HfO$_2$-terminated BaHfO$_3$ with the lattice constant reduced by 2\%.
}
\end{figure}



  
\bibliography{rashbaoxide}

\begin{thebibliography}{47}%
\makeatletter
\providecommand \@ifxundefined [1]{%
 \@ifx{#1\undefined}
}%
\providecommand \@ifnum [1]{%
 \ifnum #1\expandafter \@firstoftwo
 \else \expandafter \@secondoftwo
 \fi
}%
\providecommand \@ifx [1]{%
 \ifx #1\expandafter \@firstoftwo
 \else \expandafter \@secondoftwo
 \fi
}%
\providecommand \natexlab [1]{#1}%
\providecommand \enquote  [1]{``#1''}%
\providecommand \bibnamefont  [1]{#1}%
\providecommand \bibfnamefont [1]{#1}%
\providecommand \citenamefont [1]{#1}%
\providecommand \href@noop [0]{\@secondoftwo}%
\providecommand \href [0]{\begingroup \@sanitize@url \@href}%
\providecommand \@href[1]{\@@startlink{#1}\@@href}%
\providecommand \@@href[1]{\endgroup#1\@@endlink}%
\providecommand \@sanitize@url [0]{\catcode `\\12\catcode `\$12\catcode
  `\&12\catcode `\#12\catcode `\^12\catcode `\_12\catcode `\%12\relax}%
\providecommand \@@startlink[1]{}%
\providecommand \@@endlink[0]{}%
\providecommand \url  [0]{\begingroup\@sanitize@url \@url }%
\providecommand \@url [1]{\endgroup\@href {#1}{\urlprefix }}%
\providecommand \urlprefix  [0]{URL }%
\providecommand \Eprint [0]{\href }%
\providecommand \doibase [0]{http://dx.doi.org/}%
\providecommand \selectlanguage [0]{\@gobble}%
\providecommand \bibinfo  [0]{\@secondoftwo}%
\providecommand \bibfield  [0]{\@secondoftwo}%
\providecommand \translation [1]{[#1]}%
\providecommand \BibitemOpen [0]{}%
\providecommand \bibitemStop [0]{}%
\providecommand \bibitemNoStop [0]{.\EOS\space}%
\providecommand \EOS [0]{\spacefactor3000\relax}%
\providecommand \BibitemShut  [1]{\csname bibitem#1\endcsname}%
\let\auto@bib@innerbib\@empty
\bibitem [{\citenamefont {Ohtomo}\ and\ \citenamefont
  {Hwang}(2004)}]{Ohtomo2004}%
  \BibitemOpen
  \bibfield  {author} {\bibinfo {author} {\bibfnamefont {A.}~\bibnamefont
  {Ohtomo}}\ and\ \bibinfo {author} {\bibfnamefont {H.~Y.}\ \bibnamefont
  {Hwang}},\ }\href {\doibase 10.1038/nature02308} {\bibfield  {journal}
  {\bibinfo  {journal} {Nature}\ }\textbf {\bibinfo {volume} {427}},\ \bibinfo
  {pages} {423} (\bibinfo {year} {2004})}\BibitemShut {NoStop}%
\bibitem [{\citenamefont {Takagi}\ and\ \citenamefont
  {Hwang}(2010)}]{Takagi2010}%
  \BibitemOpen
  \bibfield  {author} {\bibinfo {author} {\bibfnamefont {H.}~\bibnamefont
  {Takagi}}\ and\ \bibinfo {author} {\bibfnamefont {H.~Y.}\ \bibnamefont
  {Hwang}},\ }\href {\doibase 10.1126/science.1182541} {\bibfield  {journal}
  {\bibinfo  {journal} {Science}\ }\textbf {\bibinfo {volume} {327}},\ \bibinfo
  {pages} {1601} (\bibinfo {year} {2010})}\BibitemShut {NoStop}%
\bibitem [{\citenamefont {Mannhart}\ and\ \citenamefont
  {Schlom}(2010)}]{Mannhart2010}%
  \BibitemOpen
  \bibfield  {author} {\bibinfo {author} {\bibfnamefont {J.}~\bibnamefont
  {Mannhart}}\ and\ \bibinfo {author} {\bibfnamefont {D.~G.}\ \bibnamefont
  {Schlom}},\ }\href {\doibase 10.1126/science.1181862} {\bibfield  {journal}
  {\bibinfo  {journal} {Science}\ }\textbf {\bibinfo {volume} {327}},\ \bibinfo
  {pages} {1607} (\bibinfo {year} {2010})}\BibitemShut {NoStop}%
\bibitem [{\citenamefont {Santander-Syro}\ \emph {et~al.}(2011)\citenamefont
  {Santander-Syro}, \citenamefont {Copie}, \citenamefont {Kondo}, \citenamefont
  {Fortuna}, \citenamefont {Pailhes}, \citenamefont {Weht}, \citenamefont
  {Qiu}, \citenamefont {Bertran}, \citenamefont {Nicolaou}, \citenamefont
  {Taleb-Ibrahimi}, \citenamefont {Le~Fevre}, \citenamefont {Herranz},
  \citenamefont {Bibes}, \citenamefont {Reyren}, \citenamefont {Apertet},
  \citenamefont {Lecoeur}, \citenamefont {Barthelemy},\ and\ \citenamefont
  {Rozenberg}}]{Santander2011}%
  \BibitemOpen
  \bibfield  {author} {\bibinfo {author} {\bibfnamefont {A.~F.}\ \bibnamefont
  {Santander-Syro}}, \bibinfo {author} {\bibfnamefont {O.}~\bibnamefont
  {Copie}}, \bibinfo {author} {\bibfnamefont {T.}~\bibnamefont {Kondo}},
  \bibinfo {author} {\bibfnamefont {F.}~\bibnamefont {Fortuna}}, \bibinfo
  {author} {\bibfnamefont {S.}~\bibnamefont {Pailhes}}, \bibinfo {author}
  {\bibfnamefont {R.}~\bibnamefont {Weht}}, \bibinfo {author} {\bibfnamefont
  {X.~G.}\ \bibnamefont {Qiu}}, \bibinfo {author} {\bibfnamefont
  {F.}~\bibnamefont {Bertran}}, \bibinfo {author} {\bibfnamefont
  {A.}~\bibnamefont {Nicolaou}}, \bibinfo {author} {\bibfnamefont
  {A.}~\bibnamefont {Taleb-Ibrahimi}}, \bibinfo {author} {\bibfnamefont
  {P.}~\bibnamefont {Le~Fevre}}, \bibinfo {author} {\bibfnamefont
  {G.}~\bibnamefont {Herranz}}, \bibinfo {author} {\bibfnamefont
  {M.}~\bibnamefont {Bibes}}, \bibinfo {author} {\bibfnamefont
  {N.}~\bibnamefont {Reyren}}, \bibinfo {author} {\bibfnamefont
  {Y.}~\bibnamefont {Apertet}}, \bibinfo {author} {\bibfnamefont
  {P.}~\bibnamefont {Lecoeur}}, \bibinfo {author} {\bibfnamefont
  {A.}~\bibnamefont {Barthelemy}}, \ and\ \bibinfo {author} {\bibfnamefont
  {M.~J.}\ \bibnamefont {Rozenberg}},\ }\href {\doibase 10.1038/nature09720}
  {\bibfield  {journal} {\bibinfo  {journal} {Nature}\ }\textbf {\bibinfo
  {volume} {469}},\ \bibinfo {pages} {189} (\bibinfo {year}
  {2011})}\BibitemShut {NoStop}%
\bibitem [{\citenamefont {Meevasana}\ \emph {et~al.}(2011)\citenamefont
  {Meevasana}, \citenamefont {King}, \citenamefont {He}, \citenamefont {Mo},
  \citenamefont {Hashimoto}, \citenamefont {Tamai}, \citenamefont
  {Songsiriritthigul}, \citenamefont {Baumberger},\ and\ \citenamefont
  {Shen}}]{Meevasana2011}%
  \BibitemOpen
  \bibfield  {author} {\bibinfo {author} {\bibfnamefont {W.}~\bibnamefont
  {Meevasana}}, \bibinfo {author} {\bibfnamefont {P.~D.~C.}\ \bibnamefont
  {King}}, \bibinfo {author} {\bibfnamefont {R.~H.}\ \bibnamefont {He}},
  \bibinfo {author} {\bibfnamefont {S.-K.}\ \bibnamefont {Mo}}, \bibinfo
  {author} {\bibfnamefont {M.}~\bibnamefont {Hashimoto}}, \bibinfo {author}
  {\bibfnamefont {A.}~\bibnamefont {Tamai}}, \bibinfo {author} {\bibfnamefont
  {P.}~\bibnamefont {Songsiriritthigul}}, \bibinfo {author} {\bibfnamefont
  {F.}~\bibnamefont {Baumberger}}, \ and\ \bibinfo {author} {\bibfnamefont
  {Z.-X.}\ \bibnamefont {Shen}},\ }\href {\doibase 10.1038/nmat2943} {\bibfield
   {journal} {\bibinfo  {journal} {Nat Mater}\ }\textbf {\bibinfo {volume}
  {10}},\ \bibinfo {pages} {114} (\bibinfo {year} {2011})}\BibitemShut
  {NoStop}%
\bibitem [{\citenamefont {Ben~Shalom}\ \emph {et~al.}(2010)\citenamefont
  {Ben~Shalom}, \citenamefont {Sachs}, \citenamefont {Rakhmilevitch},
  \citenamefont {Palevski},\ and\ \citenamefont {Dagan}}]{BenShalom2010}%
  \BibitemOpen
  \bibfield  {author} {\bibinfo {author} {\bibfnamefont {M.}~\bibnamefont
  {Ben~Shalom}}, \bibinfo {author} {\bibfnamefont {M.}~\bibnamefont {Sachs}},
  \bibinfo {author} {\bibfnamefont {D.}~\bibnamefont {Rakhmilevitch}}, \bibinfo
  {author} {\bibfnamefont {A.}~\bibnamefont {Palevski}}, \ and\ \bibinfo
  {author} {\bibfnamefont {Y.}~\bibnamefont {Dagan}},\ }\href {\doibase
  10.1103/PhysRevLett.104.126802} {\bibfield  {journal} {\bibinfo  {journal}
  {Phys. Rev. Lett.}\ }\textbf {\bibinfo {volume} {104}},\ \bibinfo {pages}
  {126802} (\bibinfo {year} {2010})}\BibitemShut {NoStop}%
\bibitem [{\citenamefont {Caviglia}\ \emph {et~al.}(2010)\citenamefont
  {Caviglia}, \citenamefont {Gabay}, \citenamefont {Gariglio}, \citenamefont
  {Reyren}, \citenamefont {Cancellieri},\ and\ \citenamefont
  {Triscone}}]{Caviglia2010}%
  \BibitemOpen
  \bibfield  {author} {\bibinfo {author} {\bibfnamefont {A.~D.}\ \bibnamefont
  {Caviglia}}, \bibinfo {author} {\bibfnamefont {M.}~\bibnamefont {Gabay}},
  \bibinfo {author} {\bibfnamefont {S.}~\bibnamefont {Gariglio}}, \bibinfo
  {author} {\bibfnamefont {N.}~\bibnamefont {Reyren}}, \bibinfo {author}
  {\bibfnamefont {C.}~\bibnamefont {Cancellieri}}, \ and\ \bibinfo {author}
  {\bibfnamefont {J.-M.}\ \bibnamefont {Triscone}},\ }\href {\doibase
  10.1103/PhysRevLett.104.126803} {\bibfield  {journal} {\bibinfo  {journal}
  {Phys. Rev. Lett.}\ }\textbf {\bibinfo {volume} {104}},\ \bibinfo {pages}
  {126803} (\bibinfo {year} {2010})}\BibitemShut {NoStop}%
\bibitem [{\citenamefont {Santander-Syro}\ \emph {et~al.}(2014)\citenamefont
  {Santander-Syro}, \citenamefont {Fortuna}, \citenamefont {Bareille},
  \citenamefont {R{\"o}del}, \citenamefont {Landolt}, \citenamefont {Plumb},
  \citenamefont {Dil},\ and\ \citenamefont {Radovi{\'c}}}]{Santander2014}%
  \BibitemOpen
  \bibfield  {author} {\bibinfo {author} {\bibfnamefont {A.~F.}\ \bibnamefont
  {Santander-Syro}}, \bibinfo {author} {\bibfnamefont {F.}~\bibnamefont
  {Fortuna}}, \bibinfo {author} {\bibfnamefont {C.}~\bibnamefont {Bareille}},
  \bibinfo {author} {\bibfnamefont {T.~C.}\ \bibnamefont {R{\"o}del}}, \bibinfo
  {author} {\bibfnamefont {G.}~\bibnamefont {Landolt}}, \bibinfo {author}
  {\bibfnamefont {N.~C.}\ \bibnamefont {Plumb}}, \bibinfo {author}
  {\bibfnamefont {J.~H.}\ \bibnamefont {Dil}}, \ and\ \bibinfo {author}
  {\bibfnamefont {M.}~\bibnamefont {Radovi{\'c}}},\ }\href {\doibase
  10.1038/nmat4107} {\bibfield  {journal} {\bibinfo  {journal} {Nat Mater}\
  }\textbf {\bibinfo {volume} {13}},\ \bibinfo {pages} {1085} (\bibinfo {year}
  {2014})}\BibitemShut {NoStop}%
\bibitem [{\citenamefont {Santander-Syro}\ \emph {et~al.}(2012)\citenamefont
  {Santander-Syro}, \citenamefont {Bareille}, \citenamefont {Fortuna},
  \citenamefont {Copie}, \citenamefont {Gabay}, \citenamefont {Bertran},
  \citenamefont {Taleb-Ibrahimi}, \citenamefont {Le~F\`evre}, \citenamefont
  {Herranz}, \citenamefont {Reyren}, \citenamefont {Bibes}, \citenamefont
  {Barth\'el\'emy}, \citenamefont {Lecoeur}, \citenamefont {Guevara},\ and\
  \citenamefont {Rozenberg}}]{Santander2012}%
  \BibitemOpen
  \bibfield  {author} {\bibinfo {author} {\bibfnamefont {A.~F.}\ \bibnamefont
  {Santander-Syro}}, \bibinfo {author} {\bibfnamefont {C.}~\bibnamefont
  {Bareille}}, \bibinfo {author} {\bibfnamefont {F.}~\bibnamefont {Fortuna}},
  \bibinfo {author} {\bibfnamefont {O.}~\bibnamefont {Copie}}, \bibinfo
  {author} {\bibfnamefont {M.}~\bibnamefont {Gabay}}, \bibinfo {author}
  {\bibfnamefont {F.}~\bibnamefont {Bertran}}, \bibinfo {author} {\bibfnamefont
  {A.}~\bibnamefont {Taleb-Ibrahimi}}, \bibinfo {author} {\bibfnamefont
  {P.}~\bibnamefont {Le~F\`evre}}, \bibinfo {author} {\bibfnamefont
  {G.}~\bibnamefont {Herranz}}, \bibinfo {author} {\bibfnamefont
  {N.}~\bibnamefont {Reyren}}, \bibinfo {author} {\bibfnamefont
  {M.}~\bibnamefont {Bibes}}, \bibinfo {author} {\bibfnamefont
  {A.}~\bibnamefont {Barth\'el\'emy}}, \bibinfo {author} {\bibfnamefont
  {P.}~\bibnamefont {Lecoeur}}, \bibinfo {author} {\bibfnamefont
  {J.}~\bibnamefont {Guevara}}, \ and\ \bibinfo {author} {\bibfnamefont
  {M.~J.}\ \bibnamefont {Rozenberg}},\ }\href {\doibase
  10.1103/PhysRevB.86.121107} {\bibfield  {journal} {\bibinfo  {journal} {Phys.
  Rev. B}\ }\textbf {\bibinfo {volume} {86}},\ \bibinfo {pages} {121107}
  (\bibinfo {year} {2012})}\BibitemShut {NoStop}%
\bibitem [{\citenamefont {King}\ \emph {et~al.}(2012)\citenamefont {King},
  \citenamefont {He}, \citenamefont {Eknapakul}, \citenamefont {Buaphet},
  \citenamefont {Mo}, \citenamefont {Kaneko}, \citenamefont {Harashima},
  \citenamefont {Hikita}, \citenamefont {Bahramy}, \citenamefont {Bell},
  \citenamefont {Hussain}, \citenamefont {Tokura}, \citenamefont {Shen},
  \citenamefont {Hwang}, \citenamefont {Baumberger},\ and\ \citenamefont
  {Meevasana}}]{King2012}%
  \BibitemOpen
  \bibfield  {author} {\bibinfo {author} {\bibfnamefont {P.~D.~C.}\
  \bibnamefont {King}}, \bibinfo {author} {\bibfnamefont {R.~H.}\ \bibnamefont
  {He}}, \bibinfo {author} {\bibfnamefont {T.}~\bibnamefont {Eknapakul}},
  \bibinfo {author} {\bibfnamefont {P.}~\bibnamefont {Buaphet}}, \bibinfo
  {author} {\bibfnamefont {S.-K.}\ \bibnamefont {Mo}}, \bibinfo {author}
  {\bibfnamefont {Y.}~\bibnamefont {Kaneko}}, \bibinfo {author} {\bibfnamefont
  {S.}~\bibnamefont {Harashima}}, \bibinfo {author} {\bibfnamefont
  {Y.}~\bibnamefont {Hikita}}, \bibinfo {author} {\bibfnamefont {M.~S.}\
  \bibnamefont {Bahramy}}, \bibinfo {author} {\bibfnamefont {C.}~\bibnamefont
  {Bell}}, \bibinfo {author} {\bibfnamefont {Z.}~\bibnamefont {Hussain}},
  \bibinfo {author} {\bibfnamefont {Y.}~\bibnamefont {Tokura}}, \bibinfo
  {author} {\bibfnamefont {Z.-X.}\ \bibnamefont {Shen}}, \bibinfo {author}
  {\bibfnamefont {H.~Y.}\ \bibnamefont {Hwang}}, \bibinfo {author}
  {\bibfnamefont {F.}~\bibnamefont {Baumberger}}, \ and\ \bibinfo {author}
  {\bibfnamefont {W.}~\bibnamefont {Meevasana}},\ }\href {\doibase
  10.1103/PhysRevLett.108.117602} {\bibfield  {journal} {\bibinfo  {journal}
  {Phys. Rev. Lett.}\ }\textbf {\bibinfo {volume} {108}},\ \bibinfo {pages}
  {117602} (\bibinfo {year} {2012})}\BibitemShut {NoStop}%
\bibitem [{\citenamefont {Reyren}\ \emph {et~al.}(2012)\citenamefont {Reyren},
  \citenamefont {Bibes}, \citenamefont {Lesne}, \citenamefont {George},
  \citenamefont {Deranlot}, \citenamefont {Collin}, \citenamefont
  {Barth\'el\'emy},\ and\ \citenamefont {Jaffr\`es}}]{Reyren2012}%
  \BibitemOpen
  \bibfield  {author} {\bibinfo {author} {\bibfnamefont {N.}~\bibnamefont
  {Reyren}}, \bibinfo {author} {\bibfnamefont {M.}~\bibnamefont {Bibes}},
  \bibinfo {author} {\bibfnamefont {E.}~\bibnamefont {Lesne}}, \bibinfo
  {author} {\bibfnamefont {J.-M.}\ \bibnamefont {George}}, \bibinfo {author}
  {\bibfnamefont {C.}~\bibnamefont {Deranlot}}, \bibinfo {author}
  {\bibfnamefont {S.}~\bibnamefont {Collin}}, \bibinfo {author} {\bibfnamefont
  {A.}~\bibnamefont {Barth\'el\'emy}}, \ and\ \bibinfo {author} {\bibfnamefont
  {H.}~\bibnamefont {Jaffr\`es}},\ }\href {\doibase
  10.1103/PhysRevLett.108.186802} {\bibfield  {journal} {\bibinfo  {journal}
  {Phys. Rev. Lett.}\ }\textbf {\bibinfo {volume} {108}},\ \bibinfo {pages}
  {186802} (\bibinfo {year} {2012})}\BibitemShut {NoStop}%
\bibitem [{\citenamefont {Kim}\ \emph {et~al.}(2012{\natexlab{a}})\citenamefont
  {Kim}, \citenamefont {Kozuka}, \citenamefont {Bell}, \citenamefont {Hikita},\
  and\ \citenamefont {Hwang}}]{MKim2012}%
  \BibitemOpen
  \bibfield  {author} {\bibinfo {author} {\bibfnamefont {M.}~\bibnamefont
  {Kim}}, \bibinfo {author} {\bibfnamefont {Y.}~\bibnamefont {Kozuka}},
  \bibinfo {author} {\bibfnamefont {C.}~\bibnamefont {Bell}}, \bibinfo {author}
  {\bibfnamefont {Y.}~\bibnamefont {Hikita}}, \ and\ \bibinfo {author}
  {\bibfnamefont {H.~Y.}\ \bibnamefont {Hwang}},\ }\href {\doibase
  10.1103/PhysRevB.86.085121} {\bibfield  {journal} {\bibinfo  {journal} {Phys.
  Rev. B}\ }\textbf {\bibinfo {volume} {86}},\ \bibinfo {pages} {085121}
  (\bibinfo {year} {2012}{\natexlab{a}})}\BibitemShut {NoStop}%
\bibitem [{\citenamefont {Bychkov}\ and\ \citenamefont
  {Rashba}(1984)}]{Bychkov1984}%
  \BibitemOpen
  \bibfield  {author} {\bibinfo {author} {\bibfnamefont {Y.~A.}\ \bibnamefont
  {Bychkov}}\ and\ \bibinfo {author} {\bibfnamefont {E.~I.}\ \bibnamefont
  {Rashba}},\ }\href@noop {} {\bibfield  {journal} {\bibinfo  {journal} {JETP
  Lett.}\ }\textbf {\bibinfo {volume} {39}},\ \bibinfo {pages} {78} (\bibinfo
  {year} {1984})}\BibitemShut {NoStop}%
\bibitem [{\citenamefont {Dresselhaus}(1955)}]{Dresselhaus1955}%
  \BibitemOpen
  \bibfield  {author} {\bibinfo {author} {\bibfnamefont {G.}~\bibnamefont
  {Dresselhaus}},\ }\href {\doibase 10.1103/PhysRev.100.580} {\bibfield
  {journal} {\bibinfo  {journal} {Phys. Rev.}\ }\textbf {\bibinfo {volume}
  {100}},\ \bibinfo {pages} {580} (\bibinfo {year} {1955})}\BibitemShut
  {NoStop}%
\bibitem [{\citenamefont {Winkler}(2003)}]{Winkler2003}%
  \BibitemOpen
  \bibfield  {author} {\bibinfo {author} {\bibfnamefont {R.}~\bibnamefont
  {Winkler}},\ }\href@noop {} {\emph {\bibinfo {title} {Spin-Orbit Coupling
  Effects in Two-Dimensional Electron and Hole Systems}}}\ (\bibinfo
  {publisher} {Springer-Verlag},\ \bibinfo {year} {2003})\BibitemShut {NoStop}%
\bibitem [{\citenamefont {Zhong}\ \emph {et~al.}(2013)\citenamefont {Zhong},
  \citenamefont {T\'oth},\ and\ \citenamefont {Held}}]{Zhong2013}%
  \BibitemOpen
  \bibfield  {author} {\bibinfo {author} {\bibfnamefont {Z.}~\bibnamefont
  {Zhong}}, \bibinfo {author} {\bibfnamefont {A.}~\bibnamefont {T\'oth}}, \
  and\ \bibinfo {author} {\bibfnamefont {K.}~\bibnamefont {Held}},\ }\href
  {\doibase 10.1103/PhysRevB.87.161102} {\bibfield  {journal} {\bibinfo
  {journal} {Phys. Rev. B}\ }\textbf {\bibinfo {volume} {87}},\ \bibinfo
  {pages} {161102} (\bibinfo {year} {2013})}\BibitemShut {NoStop}%
\bibitem [{\citenamefont {Khalsa}\ \emph {et~al.}(2013)\citenamefont {Khalsa},
  \citenamefont {Lee},\ and\ \citenamefont {MacDonald}}]{Khalsa2013}%
  \BibitemOpen
  \bibfield  {author} {\bibinfo {author} {\bibfnamefont {G.}~\bibnamefont
  {Khalsa}}, \bibinfo {author} {\bibfnamefont {B.}~\bibnamefont {Lee}}, \ and\
  \bibinfo {author} {\bibfnamefont {A.~H.}\ \bibnamefont {MacDonald}},\ }\href
  {\doibase 10.1103/PhysRevB.88.041302} {\bibfield  {journal} {\bibinfo
  {journal} {Phys. Rev. B}\ }\textbf {\bibinfo {volume} {88}},\ \bibinfo
  {pages} {041302} (\bibinfo {year} {2013})}\BibitemShut {NoStop}%
\bibitem [{\citenamefont {Kim}\ \emph {et~al.}(2014)\citenamefont {Kim},
  \citenamefont {Kang}, \citenamefont {Go},\ and\ \citenamefont
  {Han}}]{PKim2014}%
  \BibitemOpen
  \bibfield  {author} {\bibinfo {author} {\bibfnamefont {P.}~\bibnamefont
  {Kim}}, \bibinfo {author} {\bibfnamefont {K.~T.}\ \bibnamefont {Kang}},
  \bibinfo {author} {\bibfnamefont {G.}~\bibnamefont {Go}}, \ and\ \bibinfo
  {author} {\bibfnamefont {J.~H.}\ \bibnamefont {Han}},\ }\href {\doibase
  10.1103/PhysRevB.90.205423} {\bibfield  {journal} {\bibinfo  {journal} {Phys.
  Rev. B}\ }\textbf {\bibinfo {volume} {90}},\ \bibinfo {pages} {205423}
  (\bibinfo {year} {2014})}\BibitemShut {NoStop}%
\bibitem [{\citenamefont {Kim}\ \emph {et~al.}(2013)\citenamefont {Kim},
  \citenamefont {Lutchyn},\ and\ \citenamefont {Nayak}}]{YKim2013}%
  \BibitemOpen
  \bibfield  {author} {\bibinfo {author} {\bibfnamefont {Y.}~\bibnamefont
  {Kim}}, \bibinfo {author} {\bibfnamefont {R.~M.}\ \bibnamefont {Lutchyn}}, \
  and\ \bibinfo {author} {\bibfnamefont {C.}~\bibnamefont {Nayak}},\ }\href
  {\doibase 10.1103/PhysRevB.87.245121} {\bibfield  {journal} {\bibinfo
  {journal} {Phys. Rev. B}\ }\textbf {\bibinfo {volume} {87}},\ \bibinfo
  {pages} {245121} (\bibinfo {year} {2013})}\BibitemShut {NoStop}%
\bibitem [{\citenamefont {Shanavas}\ and\ \citenamefont
  {Satpathy}(2014)}]{Shanavas2014}%
  \BibitemOpen
  \bibfield  {author} {\bibinfo {author} {\bibfnamefont {K.}~\bibnamefont
  {Shanavas}}\ and\ \bibinfo {author} {\bibfnamefont {S.}~\bibnamefont
  {Satpathy}},\ }\href {\doibase 10.1103/PhysRevLett.112.086802} {\bibfield
  {journal} {\bibinfo  {journal} {Phys. Rev. Lett.}\ }\textbf {\bibinfo
  {volume} {112}},\ \bibinfo {pages} {086802} (\bibinfo {year}
  {2014})}\BibitemShut {NoStop}%
\bibitem [{\citenamefont {Park}\ \emph {et~al.}(2011)\citenamefont {Park},
  \citenamefont {Kim}, \citenamefont {Yu}, \citenamefont {Han},\ and\
  \citenamefont {Kim}}]{SPark2011}%
  \BibitemOpen
  \bibfield  {author} {\bibinfo {author} {\bibfnamefont {S.~R.}\ \bibnamefont
  {Park}}, \bibinfo {author} {\bibfnamefont {C.~H.}\ \bibnamefont {Kim}},
  \bibinfo {author} {\bibfnamefont {J.}~\bibnamefont {Yu}}, \bibinfo {author}
  {\bibfnamefont {J.~H.}\ \bibnamefont {Han}}, \ and\ \bibinfo {author}
  {\bibfnamefont {C.}~\bibnamefont {Kim}},\ }\href {\doibase
  10.1103/PhysRevLett.107.156803} {\bibfield  {journal} {\bibinfo  {journal}
  {Phys. Rev. Lett.}\ }\textbf {\bibinfo {volume} {107}},\ \bibinfo {pages}
  {156803} (\bibinfo {year} {2011})}\BibitemShut {NoStop}%
\bibitem [{\citenamefont {Kim}\ \emph {et~al.}(2012{\natexlab{b}})\citenamefont
  {Kim}, \citenamefont {Kim}, \citenamefont {Kim}, \citenamefont {Jung},
  \citenamefont {Kim}, \citenamefont {Koh}, \citenamefont {Arita},
  \citenamefont {Shimada}, \citenamefont {Namatame}, \citenamefont {Taniguchi},
  \citenamefont {Yu},\ and\ \citenamefont {Kim}}]{BKim2012}%
  \BibitemOpen
  \bibfield  {author} {\bibinfo {author} {\bibfnamefont {B.}~\bibnamefont
  {Kim}}, \bibinfo {author} {\bibfnamefont {C.~H.}\ \bibnamefont {Kim}},
  \bibinfo {author} {\bibfnamefont {P.}~\bibnamefont {Kim}}, \bibinfo {author}
  {\bibfnamefont {W.}~\bibnamefont {Jung}}, \bibinfo {author} {\bibfnamefont
  {Y.}~\bibnamefont {Kim}}, \bibinfo {author} {\bibfnamefont {Y.}~\bibnamefont
  {Koh}}, \bibinfo {author} {\bibfnamefont {M.}~\bibnamefont {Arita}}, \bibinfo
  {author} {\bibfnamefont {K.}~\bibnamefont {Shimada}}, \bibinfo {author}
  {\bibfnamefont {H.}~\bibnamefont {Namatame}}, \bibinfo {author}
  {\bibfnamefont {M.}~\bibnamefont {Taniguchi}}, \bibinfo {author}
  {\bibfnamefont {J.}~\bibnamefont {Yu}}, \ and\ \bibinfo {author}
  {\bibfnamefont {C.}~\bibnamefont {Kim}},\ }\href {\doibase
  10.1103/PhysRevB.85.195402} {\bibfield  {journal} {\bibinfo  {journal} {Phys.
  Rev. B}\ }\textbf {\bibinfo {volume} {85}},\ \bibinfo {pages} {195402}
  (\bibinfo {year} {2012}{\natexlab{b}})}\BibitemShut {NoStop}%
\bibitem [{\citenamefont {Park}\ \emph {et~al.}(2012)\citenamefont {Park},
  \citenamefont {Kim}, \citenamefont {Rhim},\ and\ \citenamefont
  {Han}}]{CPark2012}%
  \BibitemOpen
  \bibfield  {author} {\bibinfo {author} {\bibfnamefont {J.-H.}\ \bibnamefont
  {Park}}, \bibinfo {author} {\bibfnamefont {C.~H.}\ \bibnamefont {Kim}},
  \bibinfo {author} {\bibfnamefont {J.-W.}\ \bibnamefont {Rhim}}, \ and\
  \bibinfo {author} {\bibfnamefont {J.~H.}\ \bibnamefont {Han}},\ }\href
  {\doibase 10.1103/PhysRevB.85.195401} {\bibfield  {journal} {\bibinfo
  {journal} {Phys. Rev. B}\ }\textbf {\bibinfo {volume} {85}},\ \bibinfo
  {pages} {195401} (\bibinfo {year} {2012})}\BibitemShut {NoStop}%
\bibitem [{\citenamefont {Zhurova}\ \emph {et~al.}(2000)\citenamefont
  {Zhurova}, \citenamefont {Ivanov}, \citenamefont {Zavodnik},\ and\
  \citenamefont {Tsirelson}}]{Zhurova2000}%
  \BibitemOpen
  \bibfield  {author} {\bibinfo {author} {\bibfnamefont {E.~A.}\ \bibnamefont
  {Zhurova}}, \bibinfo {author} {\bibfnamefont {Y.}~\bibnamefont {Ivanov}},
  \bibinfo {author} {\bibfnamefont {V.}~\bibnamefont {Zavodnik}}, \ and\
  \bibinfo {author} {\bibfnamefont {V.}~\bibnamefont {Tsirelson}},\ }\href
  {\doibase 10.1107/S0108768100003906} {\bibfield  {journal} {\bibinfo
  {journal} {Acta Crystallogr. Sect. B-Struct.}\ }\textbf {\bibinfo {volume}
  {56}},\ \bibinfo {pages} {594} (\bibinfo {year} {2000})}\BibitemShut
  {NoStop}%
\bibitem [{\citenamefont {Maekawa}\ \emph {et~al.}(2006)\citenamefont
  {Maekawa}, \citenamefont {Kurosaki},\ and\ \citenamefont
  {Yamanaka}}]{Maekawa2006}%
  \BibitemOpen
  \bibfield  {author} {\bibinfo {author} {\bibfnamefont {T.}~\bibnamefont
  {Maekawa}}, \bibinfo {author} {\bibfnamefont {K.}~\bibnamefont {Kurosaki}}, \
  and\ \bibinfo {author} {\bibfnamefont {S.}~\bibnamefont {Yamanaka}},\ }\href
  {\doibase http://dx.doi.org/10.1016/j.jallcom.2005.06.030} {\bibfield
  {journal} {\bibinfo  {journal} {J. Alloy. Comp.}\ }\textbf {\bibinfo {volume}
  {407}},\ \bibinfo {pages} {44 } (\bibinfo {year} {2006})}\BibitemShut
  {NoStop}%
\bibitem [{\citenamefont {Scheurer}\ and\ \citenamefont
  {Schmalian}(2015)}]{Scheurer2015}%
  \BibitemOpen
  \bibfield  {author} {\bibinfo {author} {\bibfnamefont {M.~S.}\ \bibnamefont
  {Scheurer}}\ and\ \bibinfo {author} {\bibfnamefont {J.}~\bibnamefont
  {Schmalian}},\ }\href@noop {} {\bibfield  {journal} {\bibinfo  {journal} {Nat
  Commun}\ }\textbf {\bibinfo {volume} {6}},\ \bibinfo {pages} {6005} (\bibinfo
  {year} {2015})}\BibitemShut {NoStop}%
\bibitem [{\citenamefont {Shanavas}\ \emph {et~al.}(2014)\citenamefont
  {Shanavas}, \citenamefont {Popovi\ifmmode~\acute{c}\else \'{c}\fi{}},\ and\
  \citenamefont {Satpathy}}]{Shanavas2014a}%
  \BibitemOpen
  \bibfield  {author} {\bibinfo {author} {\bibfnamefont {K.~V.}\ \bibnamefont
  {Shanavas}}, \bibinfo {author} {\bibfnamefont {Z.~S.}\ \bibnamefont
  {Popovi\ifmmode~\acute{c}\else \'{c}\fi{}}}, \ and\ \bibinfo {author}
  {\bibfnamefont {S.}~\bibnamefont {Satpathy}},\ }\href {\doibase
  10.1103/PhysRevB.90.165108} {\bibfield  {journal} {\bibinfo  {journal} {Phys.
  Rev. B}\ }\textbf {\bibinfo {volume} {90}},\ \bibinfo {pages} {165108}
  (\bibinfo {year} {2014})}\BibitemShut {NoStop}%
\bibitem [{\citenamefont {{Chung}}\ \emph {et~al.}(2015)\citenamefont
  {{Chung}}, \citenamefont {{Chan}},\ and\ \citenamefont {{Yao}}}]{Chung2015}%
  \BibitemOpen
  \bibfield  {author} {\bibinfo {author} {\bibfnamefont {S.~B.}\ \bibnamefont
  {{Chung}}}, \bibinfo {author} {\bibfnamefont {C.}~\bibnamefont {{Chan}}}, \
  and\ \bibinfo {author} {\bibfnamefont {H.}~\bibnamefont {{Yao}}},\
  }\href@noop {} {\bibfield  {journal} {\bibinfo  {journal} {ArXiv e-prints}\ }
  (\bibinfo {year} {2015})},\ \Eprint {http://arxiv.org/abs/1505.00790}
  {arXiv:1505.00790 [cond-mat.supr-con]} \BibitemShut {NoStop}%
\bibitem [{\citenamefont {Schulz}(1987)}]{Schulz1987}%
  \BibitemOpen
  \bibfield  {author} {\bibinfo {author} {\bibfnamefont {H.}~\bibnamefont
  {Schulz}},\ }\href@noop {} {\bibfield  {journal} {\bibinfo  {journal}
  {Europhys. Lett.}\ }\textbf {\bibinfo {volume} {4}},\ \bibinfo {pages} {609}
  (\bibinfo {year} {1987})}\BibitemShut {NoStop}%
\bibitem [{\citenamefont {Dzyaloshinskii}(1987)}]{Dzyaloshinskii1987}%
  \BibitemOpen
  \bibfield  {author} {\bibinfo {author} {\bibfnamefont {I.}~\bibnamefont
  {Dzyaloshinskii}},\ }\href@noop {} {\bibfield  {journal} {\bibinfo  {journal}
  {JETP Lett}\ }\textbf {\bibinfo {volume} {46}},\ \bibinfo {pages} {118}
  (\bibinfo {year} {1987})}\BibitemShut {NoStop}%
\bibitem [{\citenamefont {Furukawa}\ \emph {et~al.}(1998)\citenamefont
  {Furukawa}, \citenamefont {Rice},\ and\ \citenamefont
  {Salmhofer}}]{Furukawa1998}%
  \BibitemOpen
  \bibfield  {author} {\bibinfo {author} {\bibfnamefont {N.}~\bibnamefont
  {Furukawa}}, \bibinfo {author} {\bibfnamefont {T.~M.}\ \bibnamefont {Rice}},
  \ and\ \bibinfo {author} {\bibfnamefont {M.}~\bibnamefont {Salmhofer}},\
  }\href {\doibase 10.1103/PhysRevLett.81.3195} {\bibfield  {journal} {\bibinfo
   {journal} {Phys. Rev. Lett.}\ }\textbf {\bibinfo {volume} {81}},\ \bibinfo
  {pages} {3195} (\bibinfo {year} {1998})}\BibitemShut {NoStop}%
\bibitem [{\citenamefont {Gonz\'alez}(2008)}]{Gonzalez2008}%
  \BibitemOpen
  \bibfield  {author} {\bibinfo {author} {\bibfnamefont {J.}~\bibnamefont
  {Gonz\'alez}},\ }\href {\doibase 10.1103/PhysRevB.78.205431} {\bibfield
  {journal} {\bibinfo  {journal} {Phys. Rev. B}\ }\textbf {\bibinfo {volume}
  {78}},\ \bibinfo {pages} {205431} (\bibinfo {year} {2008})}\BibitemShut
  {NoStop}%
\bibitem [{\citenamefont {Nandkishore}\ \emph {et~al.}(2012)\citenamefont
  {Nandkishore}, \citenamefont {Levitov},\ and\ \citenamefont
  {Chubukov}}]{Nandkishore2012}%
  \BibitemOpen
  \bibfield  {author} {\bibinfo {author} {\bibfnamefont {R.}~\bibnamefont
  {Nandkishore}}, \bibinfo {author} {\bibfnamefont {L.}~\bibnamefont
  {Levitov}}, \ and\ \bibinfo {author} {\bibfnamefont {A.}~\bibnamefont
  {Chubukov}},\ }\href@noop {} {\bibfield  {journal} {\bibinfo  {journal}
  {Nature Physics}\ }\textbf {\bibinfo {volume} {8}},\ \bibinfo {pages} {158}
  (\bibinfo {year} {2012})}\BibitemShut {NoStop}%
\bibitem [{\citenamefont {Meng}\ \emph {et~al.}(2015)\citenamefont {Meng},
  \citenamefont {Yang}, \citenamefont {Chen}, \citenamefont {Yao},\ and\
  \citenamefont {Kee}}]{Meng2015}%
  \BibitemOpen
  \bibfield  {author} {\bibinfo {author} {\bibfnamefont {Z.~Y.}\ \bibnamefont
  {Meng}}, \bibinfo {author} {\bibfnamefont {F.}~\bibnamefont {Yang}}, \bibinfo
  {author} {\bibfnamefont {K.-S.}\ \bibnamefont {Chen}}, \bibinfo {author}
  {\bibfnamefont {H.}~\bibnamefont {Yao}}, \ and\ \bibinfo {author}
  {\bibfnamefont {H.-Y.}\ \bibnamefont {Kee}},\ }\href {\doibase
  10.1103/PhysRevB.91.184509} {\bibfield  {journal} {\bibinfo  {journal} {Phys.
  Rev. B}\ }\textbf {\bibinfo {volume} {91}},\ \bibinfo {pages} {184509}
  (\bibinfo {year} {2015})}\BibitemShut {NoStop}%
\bibitem [{\citenamefont {Yao}\ and\ \citenamefont {Yang}(2015)}]{Yao2015}%
  \BibitemOpen
  \bibfield  {author} {\bibinfo {author} {\bibfnamefont {H.}~\bibnamefont
  {Yao}}\ and\ \bibinfo {author} {\bibfnamefont {F.}~\bibnamefont {Yang}},\
  }\href {\doibase 10.1103/PhysRevB.92.035132} {\bibfield  {journal} {\bibinfo
  {journal} {Phys. Rev. B}\ }\textbf {\bibinfo {volume} {92}},\ \bibinfo
  {pages} {035132} (\bibinfo {year} {2015})}\BibitemShut {NoStop}%
\bibitem [{\citenamefont {Chen}\ \emph {et~al.}(2015)\citenamefont {Chen},
  \citenamefont {Yao}, \citenamefont {Yao}, \citenamefont {Yang},\ and\
  \citenamefont {Ni}}]{Cheng2015}%
  \BibitemOpen
  \bibfield  {author} {\bibinfo {author} {\bibfnamefont {X.}~\bibnamefont
  {Chen}}, \bibinfo {author} {\bibfnamefont {Y.}~\bibnamefont {Yao}}, \bibinfo
  {author} {\bibfnamefont {H.}~\bibnamefont {Yao}}, \bibinfo {author}
  {\bibfnamefont {F.}~\bibnamefont {Yang}}, \ and\ \bibinfo {author}
  {\bibfnamefont {J.}~\bibnamefont {Ni}},\ }\href {\doibase
  10.1103/PhysRevB.92.174503} {\bibfield  {journal} {\bibinfo  {journal} {Phys.
  Rev. B}\ }\textbf {\bibinfo {volume} {92}},\ \bibinfo {pages} {174503}
  (\bibinfo {year} {2015})}\BibitemShut {NoStop}%
\bibitem [{\citenamefont {Yoshimatsu}\ \emph {et~al.}(2011)\citenamefont
  {Yoshimatsu}, \citenamefont {Horiba}, \citenamefont {Kumigashira},
  \citenamefont {Yoshida}, \citenamefont {Fujimori},\ and\ \citenamefont
  {Oshima}}]{Yoshimatsu2011}%
  \BibitemOpen
  \bibfield  {author} {\bibinfo {author} {\bibfnamefont {K.}~\bibnamefont
  {Yoshimatsu}}, \bibinfo {author} {\bibfnamefont {K.}~\bibnamefont {Horiba}},
  \bibinfo {author} {\bibfnamefont {H.}~\bibnamefont {Kumigashira}}, \bibinfo
  {author} {\bibfnamefont {T.}~\bibnamefont {Yoshida}}, \bibinfo {author}
  {\bibfnamefont {A.}~\bibnamefont {Fujimori}}, \ and\ \bibinfo {author}
  {\bibfnamefont {M.}~\bibnamefont {Oshima}},\ }\href {\doibase
  10.1126/science.1205771} {\bibfield  {journal} {\bibinfo  {journal}
  {Science}\ }\textbf {\bibinfo {volume} {333}},\ \bibinfo {pages} {319}
  (\bibinfo {year} {2011})}\BibitemShut {NoStop}%
\bibitem [{\citenamefont {Thiel}\ \emph {et~al.}(2006)\citenamefont {Thiel},
  \citenamefont {Hammerl}, \citenamefont {Schmehl}, \citenamefont {Schneider},\
  and\ \citenamefont {Mannhart}}]{Thiel2006}%
  \BibitemOpen
  \bibfield  {author} {\bibinfo {author} {\bibfnamefont {S.}~\bibnamefont
  {Thiel}}, \bibinfo {author} {\bibfnamefont {G.}~\bibnamefont {Hammerl}},
  \bibinfo {author} {\bibfnamefont {A.}~\bibnamefont {Schmehl}}, \bibinfo
  {author} {\bibfnamefont {C.~W.}\ \bibnamefont {Schneider}}, \ and\ \bibinfo
  {author} {\bibfnamefont {J.}~\bibnamefont {Mannhart}},\ }\href {\doibase
  10.1126/science.1131091} {\bibfield  {journal} {\bibinfo  {journal}
  {Science}\ }\textbf {\bibinfo {volume} {313}},\ \bibinfo {pages} {1942}
  (\bibinfo {year} {2006})}\BibitemShut {NoStop}%
\bibitem [{\citenamefont {Kresse}\ and\ \citenamefont
  {Hafner}(1993)}]{Kresse1993}%
  \BibitemOpen
  \bibfield  {author} {\bibinfo {author} {\bibfnamefont {G.}~\bibnamefont
  {Kresse}}\ and\ \bibinfo {author} {\bibfnamefont {J.}~\bibnamefont
  {Hafner}},\ }\href {\doibase 10.1103/PhysRevB.47.558} {\bibfield  {journal}
  {\bibinfo  {journal} {Phys. Rev. B}\ }\textbf {\bibinfo {volume} {47}},\
  \bibinfo {pages} {558} (\bibinfo {year} {1993})}\BibitemShut {NoStop}%
\bibitem [{\citenamefont {Kresse}\ and\ \citenamefont
  {Furthm\"uller}(1996)}]{Kresse1996}%
  \BibitemOpen
  \bibfield  {author} {\bibinfo {author} {\bibfnamefont {G.}~\bibnamefont
  {Kresse}}\ and\ \bibinfo {author} {\bibfnamefont {J.}~\bibnamefont
  {Furthm\"uller}},\ }\href {\doibase 10.1103/PhysRevB.54.11169} {\bibfield
  {journal} {\bibinfo  {journal} {Phys. Rev. B}\ }\textbf {\bibinfo {volume}
  {54}},\ \bibinfo {pages} {11169} (\bibinfo {year} {1996})}\BibitemShut
  {NoStop}%
\bibitem [{\citenamefont {Bl\"ochl}(1994)}]{Blochl1994}%
  \BibitemOpen
  \bibfield  {author} {\bibinfo {author} {\bibfnamefont {P.~E.}\ \bibnamefont
  {Bl\"ochl}},\ }\href {\doibase 10.1103/PhysRevB.50.17953} {\bibfield
  {journal} {\bibinfo  {journal} {Phys. Rev. B}\ }\textbf {\bibinfo {volume}
  {50}},\ \bibinfo {pages} {17953} (\bibinfo {year} {1994})}\BibitemShut
  {NoStop}%
\bibitem [{\citenamefont {Perdew}\ \emph {et~al.}(2008)\citenamefont {Perdew},
  \citenamefont {Ruzsinszky}, \citenamefont {Csonka}, \citenamefont {Vydrov},
  \citenamefont {Scuseria}, \citenamefont {Constantin}, \citenamefont {Zhou},\
  and\ \citenamefont {Burke}}]{Perdew2008}%
  \BibitemOpen
  \bibfield  {author} {\bibinfo {author} {\bibfnamefont {J.~P.}\ \bibnamefont
  {Perdew}}, \bibinfo {author} {\bibfnamefont {A.}~\bibnamefont {Ruzsinszky}},
  \bibinfo {author} {\bibfnamefont {G.~I.}\ \bibnamefont {Csonka}}, \bibinfo
  {author} {\bibfnamefont {O.~A.}\ \bibnamefont {Vydrov}}, \bibinfo {author}
  {\bibfnamefont {G.~E.}\ \bibnamefont {Scuseria}}, \bibinfo {author}
  {\bibfnamefont {L.~A.}\ \bibnamefont {Constantin}}, \bibinfo {author}
  {\bibfnamefont {X.}~\bibnamefont {Zhou}}, \ and\ \bibinfo {author}
  {\bibfnamefont {K.}~\bibnamefont {Burke}},\ }\href {\doibase
  10.1103/PhysRevLett.100.136406} {\bibfield  {journal} {\bibinfo  {journal}
  {Phys. Rev. Lett.}\ }\textbf {\bibinfo {volume} {100}},\ \bibinfo {pages}
  {136406} (\bibinfo {year} {2008})}\BibitemShut {NoStop}%
\bibitem [{\citenamefont {Marzari}\ and\ \citenamefont
  {Vanderbilt}(1997)}]{Marzari1997}%
  \BibitemOpen
  \bibfield  {author} {\bibinfo {author} {\bibfnamefont {N.}~\bibnamefont
  {Marzari}}\ and\ \bibinfo {author} {\bibfnamefont {D.}~\bibnamefont
  {Vanderbilt}},\ }\href {\doibase 10.1103/PhysRevB.56.12847} {\bibfield
  {journal} {\bibinfo  {journal} {Phys. Rev. B}\ }\textbf {\bibinfo {volume}
  {56}},\ \bibinfo {pages} {12847} (\bibinfo {year} {1997})}\BibitemShut
  {NoStop}%
\bibitem [{\citenamefont {Souza}\ \emph {et~al.}(2001)\citenamefont {Souza},
  \citenamefont {Marzari},\ and\ \citenamefont {Vanderbilt}}]{Souza2001}%
  \BibitemOpen
  \bibfield  {author} {\bibinfo {author} {\bibfnamefont {I.}~\bibnamefont
  {Souza}}, \bibinfo {author} {\bibfnamefont {N.}~\bibnamefont {Marzari}}, \
  and\ \bibinfo {author} {\bibfnamefont {D.}~\bibnamefont {Vanderbilt}},\
  }\href {\doibase 10.1103/PhysRevB.65.035109} {\bibfield  {journal} {\bibinfo
  {journal} {Phys. Rev. B}\ }\textbf {\bibinfo {volume} {65}},\ \bibinfo
  {pages} {035109} (\bibinfo {year} {2001})}\BibitemShut {NoStop}%
\bibitem [{\citenamefont {Mostofi}\ \emph {et~al.}(2008)\citenamefont
  {Mostofi}, \citenamefont {Yates}, \citenamefont {Lee}, \citenamefont {Souza},
  \citenamefont {Vanderbilt},\ and\ \citenamefont {Marzari}}]{Mostofi2008}%
  \BibitemOpen
  \bibfield  {author} {\bibinfo {author} {\bibfnamefont {A.~A.}\ \bibnamefont
  {Mostofi}}, \bibinfo {author} {\bibfnamefont {J.~R.}\ \bibnamefont {Yates}},
  \bibinfo {author} {\bibfnamefont {Y.-S.}\ \bibnamefont {Lee}}, \bibinfo
  {author} {\bibfnamefont {I.}~\bibnamefont {Souza}}, \bibinfo {author}
  {\bibfnamefont {D.}~\bibnamefont {Vanderbilt}}, \ and\ \bibinfo {author}
  {\bibfnamefont {N.}~\bibnamefont {Marzari}},\ }\href {\doibase
  http://dx.doi.org/10.1016/j.cpc.2007.11.016} {\bibfield  {journal} {\bibinfo
  {journal} {Computer Physics Communications}\ }\textbf {\bibinfo {volume}
  {178}},\ \bibinfo {pages} {685 } (\bibinfo {year} {2008})}\BibitemShut
  {NoStop}%
\bibitem [{\citenamefont {Stroppa}\ \emph {et~al.}(2014)\citenamefont
  {Stroppa}, \citenamefont {Di~Sante}, \citenamefont {Barone}, \citenamefont
  {Bokdam}, \citenamefont {Kresse}, \citenamefont {Franchini}, \citenamefont
  {Whangbo},\ and\ \citenamefont {Picozzi}}]{Stroppa2014}%
  \BibitemOpen
  \bibfield  {author} {\bibinfo {author} {\bibfnamefont {A.}~\bibnamefont
  {Stroppa}}, \bibinfo {author} {\bibfnamefont {D.}~\bibnamefont {Di~Sante}},
  \bibinfo {author} {\bibfnamefont {P.}~\bibnamefont {Barone}}, \bibinfo
  {author} {\bibfnamefont {M.}~\bibnamefont {Bokdam}}, \bibinfo {author}
  {\bibfnamefont {G.}~\bibnamefont {Kresse}}, \bibinfo {author} {\bibfnamefont
  {C.}~\bibnamefont {Franchini}}, \bibinfo {author} {\bibfnamefont {M.-H.}\
  \bibnamefont {Whangbo}}, \ and\ \bibinfo {author} {\bibfnamefont
  {S.}~\bibnamefont {Picozzi}},\ }\href@noop {} {\bibfield  {journal} {\bibinfo
   {journal} {Nat Commun}\ }\textbf {\bibinfo {volume} {5}},\ \bibinfo {pages}
  {5900} (\bibinfo {year} {2014})}\BibitemShut {NoStop}%
\bibitem [{\citenamefont {Yu}\ and\ \citenamefont {Cardona}(2010)}]{Yu2010}%
  \BibitemOpen
  \bibfield  {author} {\bibinfo {author} {\bibfnamefont {P.}~\bibnamefont
  {Yu}}\ and\ \bibinfo {author} {\bibfnamefont {M.}~\bibnamefont {Cardona}},\
  }\href@noop {} {\emph {\bibinfo {title} {Fundamentals of Semiconductors:
  Physics and Materials Properties}}}\ (\bibinfo  {publisher}
  {Springer-Verlag},\ \bibinfo {year} {2010})\BibitemShut {NoStop}%
\end{thebibliography}%

\end{document}